%% file: main.tex
\documentclass[aps,prd,preprintnumbers,amsmath,amssymb,nofootinbib,superscriptaddress,floatfix,showkeys,twocolumn]{revtex4-1}
\usepackage[utf8]{inputenc}
\usepackage{graphicx,color}
\usepackage[dvipsnames]{xcolor}
\usepackage{amsmath,amssymb}
\usepackage{hyperref}
\usepackage{epstopdf}
\usepackage{slashed}
\usepackage{soul}
\usepackage{bm}
\usepackage{amssymb}
\usepackage[normalem]{ulem}
\usepackage[caption=false]{subfig}
\usepackage{changes}
\usepackage{booktabs}

\newcommand{\lsim}{\mathrel{\mathop{\kern 0pt \rlap
  {\raise.2ex\hbox{$<$}}}
  \lower.9ex\hbox{\kern-.190em $\sim$}}}
\newcommand{\gsim}{\mathrel{\mathop{\kern 0pt \rlap
  {\raise.2ex\hbox{$>$}}}
  \lower.9ex\hbox{\kern-.190em $\sim$}}}

\DeclareUnicodeCharacter{2212}{\ensuremath{-}}

\definechangesauthor[name={Cui}, color=blue]{Cui}
\definechangesauthor[name={yyli}, color=purple]{yyli}

\input{ads_macros}

\begin{document}

\title{The impact of primordial magnetic fields on the formation of galaxies}
\author{Yao-Yu Li}
\email{yyli@pmo.ac.cn}
\affiliation{Key Laboratory of Dark Matter and Space Astronomy,
Purple Mountain Observatory, Chinese Academy of Sciences, Nanjing 210023, China}
\affiliation{School of Astronomy and Space Science, University of Science and Technology of China, Hefei 230026, China}

\author{Shihong Liao}
\email{shliao@nao.cas.cn}
\affiliation{National Astronomical Observatories, Chinese Academy of Sciences, 20A Datun Road, Chaoyang District, Beijing 100101, China}
\affiliation{School of Astronomy and Space Science, University of Chinese Academy of Sciences, Beijing 100049, China}

\author{Yue-Lin Sming Tsai}
\affiliation{Key Laboratory of Dark Matter and Space Astronomy,
Purple Mountain Observatory, Chinese Academy of Sciences, Nanjing 210023, China}
\affiliation{School of Astronomy and Space Science, University of Science and Technology of China, Hefei 230026, China}

\author{Yi-Zhong Fan}
\affiliation{Key Laboratory of Dark Matter and Space Astronomy,
Purple Mountain Observatory, Chinese Academy of Sciences, Nanjing 210023, China}
\affiliation{School of Astronomy and Space Science, University of Science and Technology of China, Hefei 230026, China}

\begin{abstract}
Primordial magnetic fields (PMFs), potentially generated during cosmic inflation or early-universe phase transitions, can modify cosmic structure formation by enhancing the post-recombination matter power spectrum. In this work, we investigate the impact of PMFs on galaxy formation using a suite of high-resolution cosmological hydrodynamical simulations evolved to $z=0$. Our simulations are initialized with PMF-induced enhancements to the matter power spectrum and include more comprehensive baryonic subgrid models than previous work, incorporating both stellar and AGN feedback. We find that PMFs systematically accelerate early structure formation, producing more abundant halos and galaxies, higher baryon fractions, enhanced star formation rates, and rapid growth of supermassive black holes at high redshifts ($z \sim 10$). However, the resulting enhanced stellar and AGN feedback efficiently processes and expels gas, reducing the late-time differences between PMF and standard $\Lambda$CDM models. The PMF signatures are therefore strongest at high redshift and gradually weaken toward low redshift as nonlinear evolution and baryonic feedback dominate over the initial PMF-induced enhancements. The impact of PMFs depends strongly on the characteristic peak scale, $k_{\rm peak}$, of the initial power enhancement, with models yielding similar non-linear evolution if they share a similar $k_{\rm peak}$, even when other PMF parameters differ. Our results highlight the critical role of comprehensive baryonic physics in accurately quantifying PMF signatures.
 
\end{abstract}

\maketitle
\section{Introduction}
Magnetic fields are ubiquitous in the observable Universe — from the planets to stars, galaxies, clusters and the cosmic web. However, the origin of these magnetic fields remains largely unknown. One compelling possibility is that weak seed fields were produced in the early Universe and later amplified during structure formation, producing the complex field geometries we detect today \cite{Subramanian:2015lua, Banerjee:2004df, 2013A&ARv..21...62D, Brandenburg:1996fc}. Gamma-ray observations of TeV blazars and gamma-ray bursts have been used to probe weak intergalactic magnetic fields through their effects on electromagnetic cascades \cite{Xia:2022uua, 2010Sci...328...73N}. In particular, the absence of an expected secondary GeV extension in some blazar spectra, which arises from the inverse Compton scattering of cosmic microwave background (CMB) photons by relativistic electron-positron pairs created in the cascade, suggests the presence of intergalactic magnetic fields exceeding $\sim 10^{-16}$ G on Mpc scales \cite{2010Sci...328...73N,2010MNRAS.406L..70T,Vachaspati:2020blt}.\footnote{Note that an alternative explanation for the deficit of secondary GeV emission is plasma instabilities that dissipate the pair beam energy \cite{Broderick:2018nqf}.} Because cosmic voids lack significant local astrophysical sources such as supernovae or galactic dynamos, these void fields are widely interpreted as primordial magnetic fields (PMFs).
Such fields may have been generated during cosmic inflation \cite{PhysRevD.37.2743, 1992ApJ...391L...1R, 2005PhRvD..71j3509A, 2018PhRvD..97h3503S} or early-Universe phase transitions (e.g., the QCD or electroweak transitions) \cite{1989ApJ...344L..49Q, 1991PhLB..265..258V, 1997PhRvD..55.4582S, 2008PhRvL.100x1301D, 2017PhRvD..96l3528B, 2019PhRvD.100h3006Z,Di:2025ncl,Yang:2021uid} , and would therefore provide a unique window onto high-energy physics. Moreover, PMFs persist and evolve with the expanding Universe, imprinting signatures across multiple epochs from recombination \cite{Jedamzik:2020krr,Jedamzik:2018itu, 2024PhRvD.109d3538L} to the formation of the first stars \cite{Koh:2021gjt,Zhang:2024yph} and through later stages of structure formation \cite{Ralegankar:2023pyx, Tashiro:2005ua, Cruz:2023rmo,Kahniashvili:2012dy}. This combination of observable imprints offers a strong observational- and simulation-driven impetus to study PMFs.

Observational efforts over the past decade have placed increasingly stringent constraints on the properties of PMFs. Measurements of the CMB
anisotropies and polarization have yielded some of the most robust upper limits on the amplitude of PMFs. For example, analyses of the {\it Planck} temperature and polarization power spectra constrain the comoving field strength to be $B_{\rm 1~Mpc} \lesssim \mathcal{O}(1)\,{\rm nG}$ at 1 Mpc scales for nearly scale‑invariant spectra, with even tighter limits when including non‑Gaussianity and Faraday rotation measurements \cite{Planck:2015zrl, Zucca:2016iur, Trivedi:2013wqa, Pshirkov:2015tua}. The upcoming {\it LiteBIRD} satellite is expected to significantly improve constraints on the properties of PMFs \cite{LiteBIRD:2024twk}.

Beyond their imprints on the CMB, PMFs significantly impact the formation of structures by enhancing matter fluctuations on small scales. These modifications can be targeted by probes such as weak lensing and line-intensity mapping, with forecast sensitivities reaching ${\sim}0.1$ nG for the former and 0.006–1 nG for the latter \cite{Fedeli:2012rr,Adi:2023qdf}. The Ly$\alpha$ forest serves as another highly sensitive diagnostic. Because its flux power spectrum responds strongly to PMF-induced perturbations in the linear matter field, recent analyses have placed a rigorous bound of $B_{\rm 1\,Mpc} < 0.30\,{\rm nG}$ for $n_B = -2.9$ \cite{Pavicevic:2025gqi}. Furthermore, by altering the thermal and ionization evolution of the intergalactic medium, these small-scale fluctuations can dramatically reshape the redshifted 21-cm signal, with recent forecasts indicating that interferometers like HERA can isolate PMF signatures down to the pG level \cite{Cruz:2023rmo}. Dwarf galaxies offer a complementary probe for constraining PMFs, and recent studies show that moderate fields ($B_{\rm 1~Mpc} \lesssim 0.10\,{\rm nG}$) successfully reproduce the observed scaling relations of Local Group dwarfs \cite{Sanati:2020oay, Sanati:2024ijt}.

The robustness of PMF constraints from structure formation depends on accurately quantifying their impact on structure formation. This, in turn, requires high-resolution cosmological simulations capable of resolving the complicated galaxy formation processes within the non-linear evolutionary regime. A growing body of numerical work has been dedicated to this area, yielding significant progress in recent years. For instance, cosmological zoom-in hydrodynamical simulations by \cite{Sanati:2020oay, Sanati:2024ijt} demonstrated that PMFs critically influence the line-of-sight velocity dispersion, half-light radius, metallicity, and luminosity of classical and ultra-faint dwarf galaxies at $z=0$, providing a pathway to constrain PMFs using Local Group observations. Using the SPHINX-MHD radiation-magnetohydrodynamical simulations evolved down to $z = 6$, \citet{Katz:2021iou} showed that strong PMFs accelerate early halo collapse and can produce reionization histories inconsistent with CMB data, while identifying the global 21-cm signal and reduced galaxy sizes as key PMF signatures. Utilizing cosmological simulations accounting for the distinct impact of PMFs on baryon and dark matter perturbations in the initial conditions (ICs), \citet{Ralegankar:2024ekl} found that PMFs preferentially enhance baryon density perturbations, creating high-redshift ($z \gtrsim 4$) galaxies with baryon fractions significantly above the cosmic average and boosted star formation rates. Most recently, \citet{Pavicevic:2025gqi} performed a systematic investigation of PMF effects on the Ly$\alpha$ forest.

Despite these advancements, several fundamental challenges and limitations remain. Some of the aforementioned simulations are limited by small box sizes (e.g., $L_{\rm box} = 5$~Mpc in \cite{Katz:2021iou}) or restricted to high redshifts (e.g., $z >6$ in \cite{Katz:2021iou} and some runs with $z > 4$ in \cite{Ralegankar:2024ekl}), or employ zoom-in techniques that lack the statistical power of large-volume samples (e.g., \cite{Sanati:2020oay, Sanati:2024ijt}). A more fundamental challenge lies in the potential degeneracy between PMF-induced signals and baryonic processes. As there is currently no theoretical consensus on the optimal numerical implementation of subgrid baryonic physics (see e.g., \cite{Naab:2017araa, Vogelsberger:2019ynw, Crain:2023xap}), independent simulations utilizing diverse modeling frameworks are essential to verify the robustness of PMF constraints. Notably, previous simulations have largely omitted AGN feedback, a critical mechanism that influences the evolution of massive galaxies and their surrounding environments.

In this work, we perform a series of high-resolution cosmological hydrodynamical simulations (with $L_{\rm box} = 25$ Mpc, run down to $z = 0$) using a modified version of the \textsc{Gadget-3} code \cite{Springel:2005mi}. Our setup incorporates PMF effects in the IC and includes a comprehensive suite of baryonic processes, notably both stellar and AGN feedback. These simulations allow us to investigate the impact of PMFs on structure formation across a broad range of scales and redshifts, and the role of baryonic feedback processes in complicating the resulting signatures.

This paper is organized as follows. In Section \ref{sec:theory}, we outline the theoretical impact of PMFs on density perturbations and matter power spectra. Section \ref{sec:sim} describes the simulation methodology, including the numerical code, the adopted PMF models, and the construction of ICs. 
In Section \ref{sec:res}, we present our main results. We first examine the impact of PMFs on the matter distribution and the evolution of the power spectra, and then analyze their effects on halo and galaxy properties. 
Finally, we summarize and discuss our main findings in Section \ref{sec:summary}.

\section{Impact of Primordial Magnetic Fields on density perturbations} 
\label{sec:theory}
In this section, we briefly review how PMFs influence the evolution of cosmological density perturbations. We consider the PMFs generated during the inflationary epoch (with spectral index $n_B<0$), but concentrate on their effects in the post-recombination Universe. Prior to recombination, the tight coupling between photons and baryons leads to strong photon pressure and viscous damping that suppress the small-scale perturbations, preventing PMFs from leaving a substantial, long-lived imprint on baryon inhomogeneities. 

After recombination, however, baryons rapidly decouple from photons and the photon drag term becomes negligible. The Lorentz force from the PMFs can directly induce motions in the ionized baryons. This motion is in turn transferred to the neutral components via diople interactions, thereby driving the entire baryonic fluid and ultimately generating baryon density fluctuations. Because baryons and dark matter are coupled through gravitational force, these PMF-driven baryon motions also source dark matter perturbations and thus enhance the total matter power spectrum on small scales \cite{2005MNRAS.356..778S,2004PhRvD..70l3003B,2005MNRAS.356..778S}.

\subsection{Linear density evolution in the presence of PMFs}
In addition to being accelerated by the gravitational potential, baryons are directly affected by PMFs through the Lorentz force, as described above. In this study, we focus on scales much larger than the thermal Jeans length, so the role of baryon thermal pressure can be neglected. In comoving coordinates the baryon Euler equation with the magnetic contribution reads \cite{2004PhRvD..70l3003B,Subramanian:2015lua,2013A&ARv..21...62D}
\begin{align}
\frac{\partial \boldsymbol{v}_{\rm b}(\boldsymbol{x},t)}{\partial t} + H(t)\boldsymbol{v}_{\rm b}(\boldsymbol{x},t) 
&= -\frac{\nabla\psi(\boldsymbol{x},t)}{a(t)}  \nonumber \\
& \quad + \frac{[\nabla\times\boldsymbol{B}(\boldsymbol{x},t)]\times\boldsymbol{B}(\boldsymbol{x},t)}{4\pi a(t)\rho_{\rm b}(t)},
\end{align}
where $\boldsymbol{v}_{\rm b}(\boldsymbol{x},t)$ denotes the peculiar velocity of the baryon fluid, and $H(t) \equiv \dot{a}(t)/a(t)$ represents the Hubble expansion rate, with $a(t)$ being the cosmic scale factor. The gravitational influence is captured by the scalar potential $\psi(\boldsymbol{x},t)$. The second term on the right-hand side describes the Lorentz force exerted on the baryons, where $\boldsymbol{B}(\boldsymbol{x},t)$ is the physical magnetic field and $\rho_{\rm b}(t)$ is the background baryon energy density. 

Under the ideal magnetohydrodynamic (MHD) assumption (infinite conductivity, no battery terms), the induction equation can be written as \cite{2004PhRvD..70l3003B,Subramanian:2015lua,2013A&ARv..21...62D} 
\begin{equation}
\frac{\partial}{\partial t}\left[a^2(t)\boldsymbol{B}(\boldsymbol{x},t)\right] = \frac{\nabla\times\left[\boldsymbol{v}_{\rm b}(\boldsymbol{x},t)\times a^2(t)\boldsymbol{B}(\boldsymbol{x},t)\right]}{a(t)}.
\end{equation}

On sufficiently large scales, the Alfvén crossing time exceeds the Hubble time, rendering magnetic backreaction on linear, large-scale baryon dynamics negligible. Therefore, the induction term can be dropped in the linear regime and the magnetic field is, to leading order, frozen into the cosmological expansion:
\begin{equation}
a^2(t)\boldsymbol{B}(\boldsymbol{x},t) = \boldsymbol{B}(\boldsymbol{x},t_0) \equiv \boldsymbol{B}_0(\boldsymbol{x}),
\end{equation}
where $t_0$ denotes the present cosmic time, and the scale factor is normalized as $a(t_0)=1$.

Since the continuity and Poisson equations retain their standard forms, the evolution equation for the baryon density perturbation follows \cite{1980lssu.book.....P,1996ApJ...468...28K,2023PhRvD.108b3521A}
\begin{align}
\frac{\partial^2 \delta_{\rm b}(\boldsymbol{x},t)}{\partial t^2}
&= -2H(t)\frac{\partial \delta_{\rm b}(\boldsymbol{x},t)}{\partial t}+\frac{\boldsymbol{S}_0(\boldsymbol{x})}{a^3(t)} \nonumber\\
&\quad + 4\pi G\bigl[\rho_{\rm DM}(t)\delta_{\rm DM}(\boldsymbol{x},t) + \rho_{\rm b}(t)\delta_{\rm b}(\boldsymbol{x},t)\bigr],
\label{eq:delta_b_t}
\end{align}
where $\delta_{b}(\boldsymbol{x},t)$ and $\delta_{\rm DM}(\boldsymbol{x},t)$ correspond to the density fluctuations of baryons and dark matter, $\rho_{\rm DM}(t)$ is the background dark matter energy density and $\boldsymbol{S}_0(\boldsymbol{x})$ is the source term from magnetic fields:
\begin{equation}
\boldsymbol{S}_0(\boldsymbol{x})=
\frac{\nabla\cdot\left[(\nabla\times\boldsymbol{B}_0(\boldsymbol{x})\times\boldsymbol{B}_0(\boldsymbol{x})\right]}{4\pi\rho_{b,0}}=\text{constant}.
\end{equation}
Here $\rho_{b,0} = \rho_b a^{3}$ is the present-day value of baryon energy density. 

In contrast to baryons, dark matter does not couple directly to magnetic fields. Therefore, its density perturbations evolve according to the standard gravitational-instability equation \cite{1980lssu.book.....P,1996ApJ...468...28K,2023PhRvD.108b3521A}:
\begin{align}
\frac{\partial^2 \delta_{\rm DM}(\boldsymbol{x},t)}{\partial t^2} 
&= -2H(t)\frac{\partial \delta_{\rm DM}(\boldsymbol{x},t)}{\partial t} \nonumber\\
&\quad + 4\pi G\bigl[\rho_{\rm DM}(t)\delta_{\rm DM}(\boldsymbol{x},t) + \rho_b(t)\delta_b(\boldsymbol{x},t)\bigr].
\label{eq:delta_DM_t}
\end{align}

While most previous studies have combined Eq.~\eqref{eq:delta_b_t} and Eq.~\eqref{eq:delta_DM_t} to focus solely on the evolution of the total matter density perturbation, this work takes a different approach \cite{Ralegankar:2024ekl}. Recognizing that PMFs impact baryons and dark matter to different extents, we solve these two equations simultaneously. For computational convenience, we perform a change of variables from cosmic time $t$ to the scale factor $a$ to recast the evolution equations. Utilizing the definition of the critical density, we have $4\pi G \rho_{\rm DM,0} = \frac{3}{2} H_0^2 \Omega_{\rm DM,0}$ and $4\pi G \rho_{\rm b,0} = \frac{3}{2} H_0^2 \Omega_{\rm b,0}$. During the matter-dominated epoch (accounting for the transition from the radiation-dominated era), the Hubble parameter satisfies:$H^2(a) = \frac{8\pi G}{3} \rho_{\rm m,0} a^{-3} \left[ 1 + \frac{a_{\rm eq}}{a} \right] = H_0^2 \Omega_{\rm m,0} a^{-3} \left( 1 + \frac{a_{\rm eq}}{a} \right)$. Here, $H_0$ is the Hubble constant, while $\Omega_{\rm DM,0}$, $\Omega_{\rm b,0}$, and $\Omega_{\rm m,0}$ represent the present-day density parameters for dark matter, baryons, and total matter, respectively. The terms $\rho_{\rm DM,0}$ and $\rho_{\rm m,0}$ denote the corresponding current energy densities, and $a_{\rm eq}$ signifies the scale factor at the epoch of radiation-matter equality. Combining the above relations, we arrive at the following evolution equations:

\begin{align}
&a^2\frac{\partial^2\delta_{\rm b}(\boldsymbol{x},a)}{\partial a^2} + a\frac{3}{2}\frac{\partial\delta_{\rm b}(\boldsymbol{x},a)}{\partial a}  \nonumber\\
&\quad = \frac{3}{2}\frac{\Omega_{\rm b,0}\delta_{\rm b}(\boldsymbol{x},a) + \Omega_{\rm DM,0}\delta_{\rm DM}(\boldsymbol{x},a)}{\Omega_{\rm m,0}(1+a_{\rm eq}/a)} \nonumber \\
&\quad \quad +\frac{\boldsymbol{S}_0(\boldsymbol{x})}{H_0^2 \Omega_{\rm m,0}(1+a_{\rm eq}/a)}, \label{eq:delta_b_a}\\
&a^2\frac{\partial^2\delta_{\rm DM}(\boldsymbol{x},a)}{\partial a^2} + a\frac{3}{2}\frac{\partial\delta_{\rm DM}(\boldsymbol{x},a)}{\partial a}  \nonumber\\
&\quad = \frac{3}{2}\frac{\Omega_{\rm DM,0}\delta_{\rm DM}(\boldsymbol{x},a) + \Omega_{\rm b,0}\delta_{\rm b}(\boldsymbol{x},a)}{\Omega_{\rm m,0}(1+a_{\rm eq}/a)}. \label{eq:delta_DM_a}
\end{align}

The general solution can be expressed as a linear combination of two components: $\delta = \delta^{\Lambda\text{CDM}} + \delta^{\text{PMF}}$. Here, $\delta^{\Lambda\text{CDM}}$ describes the standard evolution from primordial curvature perturbations, while $\delta^{\text{PMF}}$ captures the inhomogeneous response driven by the magnetic source $\boldsymbol{S}_0(\boldsymbol{x})$. Linearity of the governing equations justifies this decomposition. Since $\delta^{\Lambda\text{CDM}}$ is well studied, we focus on the magnetically-induced term. 

As mentioned earlier, the high opacity of the pre-recombination plasma and strong Compton scattering effectively “freeze” the baryon fluid, preventing the magnetic field from generating significant fluctuations. Therefore, it is reasonable to set the initial conditions for the PMF-induced perturbations at recombination ($a = a_{\rm rec}$) as $\delta_{\text{b}}^{\text{PMF}} = \delta_{\text{DM}}^{\text{PMF}} = 0$ and $\partial\delta_{\text{b}}^{\text{PMF}}/\partial a = \partial\delta_{\text{DM}}^{\text{PMF}}/\partial a  = 0$. After recombination, when photon drag becomes negligible, the evolution of $\delta^{\rm PMF}$ is governed solely by the magnetic source $\boldsymbol{S}_0(\boldsymbol{x})$. The post-recombination density perturbations can be expressed as 
\begin{align}
&\delta_{\text{b}}^{\text{PMF}}(\boldsymbol{x},a) = \xi_{\text{b}}(a)\frac{\boldsymbol{S}_0(\boldsymbol{x})}{H_0^2 \Omega_{\rm m,0}(1+a_{\rm eq}/a)},  \nonumber\\ &\delta_{\text{DM}}^{\text{PMF}}(\boldsymbol{x},a) = \xi_{\text{DM}}(a)\frac{\boldsymbol{S}_0(\boldsymbol{x})}{H_0^2 \Omega_{\rm m,0}(1+a_{\rm eq}/a)}.
\label{eq:delta_pmf}
\end{align}

The time dependence $\xi_{\rm b}(a)$ and $\xi_{\rm DM}(a)$ can be easily solved by substituting Eq.~\eqref{eq:delta_pmf} into Eq.~\eqref{eq:delta_b_a} and Eq.~\eqref{eq:delta_DM_a}, which effectively cancels out the $\boldsymbol{S}_0(\boldsymbol{x})$ term.

\subsection{Impact of PMFs on the matter power spectrum}
As indicated by Eq. \eqref{eq:delta_pmf}, in order to determine the impact of PMFs on the baryon and dark matter power spectra, the most crucial step is to evaluate the power spectrum of $\boldsymbol{S}_0(\boldsymbol{x})$. In this study, we assume that the PMFs are non-helical\footnote{If a helical component were included, the magnetic two-point correlation function in Eq. \eqref{two-point correlation} would contain an additional term. Moreover, the approximate conservation of magnetic helicity during MHD evolution can drive an inverse transfer of magnetic energy to larger scales, resulting in a slower decay of magnetic energy and a faster growth of the magnetic correlation length than in the non-helical case \cite{2004PhRvD..70l3003B,Kahniashvili:2010gp,Ballardini:2014jta}.}, allowing us to define their power spectrum $P_{\rm B}(k)$ as follows:
\begin{equation}
\langle \tilde{B}_i(\boldsymbol{k})\tilde{B}_j^*(\boldsymbol{k}')\rangle = (2\pi)^3\delta_D\!\left(\boldsymbol{k}-\boldsymbol{k}'\right)\frac{P_{ij}}{2}P_B(k),
\label{two-point correlation}
\end{equation}
where $\tilde{B}(\boldsymbol{k})=\int B(\boldsymbol{x})\,e^{-i\boldsymbol{k}\cdot\boldsymbol{x}}\,d^3x$ is the Fourier transform of $B(\boldsymbol{x})$, and $P_{ij}=\left(\delta_{ij}-k_i k_j/k^2\right)$ is the projection tensor. Furthermore, we adopt a power-law form for the magnetic power spectrum $P_{\rm B}(k)$, which is modulated by an exponential cut-off to account for the turbulent damping effects at small scale \cite{1998PhRvD..58h3502S,2012PhRvD..86d3510S,Ralegankar:2024ekl} : 
\begin{equation}
P_{\rm B}(k) = A_{\rm B} k^{n_{\rm B}} \exp\left( -{k^2}{\lambda_{\rm D}^2} \right).
\label{pmf pk}
\end{equation}
In this expression, $n_{\rm B}$ denotes the magnetic spectral index. To ensure a finite magnetic energy density on large scales, and to avoid the large theoretical uncertainties that arise when the power spectrum becomes highly sensitive to the small-scale cut-off $\lambda_{\rm D}$ for steeper spectra, we restrict our study to models with $-3 < n_{\rm B} < -1.5$\footnote{A detailed study of models with $n_{\rm B} > -1.5$, especially those generated causally after inflation and therefore subject to the horizon-scale constraint on their initial correlation length at magnetogenesis, requires full MHD simulations \cite{Ralegankar:2024arh,Durrer:2003ja,PhysRevD.96.123528,Kahniashvili:2012uj}. Besides, the exponential cut-off captures the suppression of small-scale PMF modes by MHD turbulence, but does not follow the full time-dependent decay and redistribution of magnetic power discussed in previous MHD studies \cite{Brandenburg:2024tyi,Kahniashvili:2010gp,Jedamzik:2023rfd,Trivedi:2018ejz}. Since we focus on scales larger than $\lambda_{\rm D}$, whose evolution is insensitive to the detailed damping behavior below $\lambda_{\rm D}$, these effects are not included in the present work.}. The amplitude $A_{\rm B}$ is related to the effective comoving magnetic field strength $B_{\lambda}$, obtained by smoothing the field with a Gaussian filter over a characteristic scale $\lambda$ (in this study, we set $\lambda = 1 \text{ Mpc}$): 
\begin{equation}
B_{\lambda}^2 = \int_0^\infty \frac{dk k^2}{2\pi^2} P_B(k) e^{-k^2\lambda^2}. 
\end{equation}
By performing this smoothing, $A_{\rm B}(t)$ can be expressed as
\begin{equation}
A_{\rm B} = \frac{(2\pi)^2 B_{\lambda}^2}{\Gamma\left(\frac{n_{\rm B}+3}{2}\right)} \lambda^{n_{\rm B}+3},
\end{equation}
where $\Gamma(x)$ is the Gamma function. 

The exponential term in Eq. \eqref{pmf pk} characterizes the dissipation of magnetic energy into the fluid via MHD turbulence, arising from the back-reaction of baryons on PMFs below the damping scale \cite{Ralegankar:2024ekl}
\begin{equation}
\lambda_{\rm D} = \left[0.1 \times \kappa_{n_{\rm B}} \times \left(\frac{B_{\lambda}}{1\,{\rm nG}}\right)\right]^{2/(n_B+5)} {\rm Mpc}.
\label{lambda_D}
\end{equation}
Here, the coefficient $\kappa_{n_{\rm B}}$ is determined by the magnetic spectral index $n_{\rm B}$ and governs the amplitude of the PMF-induced matter power spectrum. We will provide a more detailed discussion later.

In addition to the damping scale, another characteristic scale frequently discussed in previous literature is the magnetic Jeans scale, $\lambda_{\rm J}$. Conventionally, $\lambda_{\rm J}$ is defined by identifying the scale at which magnetic pressure balances gravitational collapse, analogous to the classic thermal Jeans scale but with the sound speed replaced by the Alfvén velocity $v_A$. In this work, we follow the approach of \cite{Ralegankar:2024ekl} and adopt $\lambda_{\rm D}$ as our primary characteristic scale instead of $\lambda_{\rm J}$. While $\lambda_{\rm J}$ is often framed within a static balance narrative, it is conceptually incomplete for PMFs, which actively source the initial growth of baryon perturbations. The suppression of power is more physically described by $\lambda_{\rm D}$, which marks the threshold where MHD turbulence dissipates these perturbations before gravity can lead to non-linear collapse. On scales larger than $\lambda_{\rm D}$, gravity eventually dominates the Lorentz force, ensuring that $\lambda_{\rm D}$ effectively fulfills the same physical role as the magnetic Jeans scale in defining the small-scale cut-off of the power spectrum.

Utilizing the PMF power spectrum, we evaluate the Fourier transform of the Lorentz force source term, $\tilde{\boldsymbol{S}}_0(\boldsymbol{k})$. The resulting baryon power spectrum induced by PMFs, $P_{\rm b}^{\rm PMF}(k)$, is then derived from the two-point correlation function of density perturbations, $\langle \tilde{\delta}_b(\boldsymbol{k})\tilde{\delta}_b^*(\boldsymbol{k}')\rangle$ \cite{Adi:2023doe}. Following the formalism in \cite{Ralegankar:2024ekl}, the dimensionless power spectrum of baryon is given by:
\begin{align}
\Delta_{\rm b}^{\rm PMF}(k) 
&\equiv \frac{k^3 P_{\rm b}^{\rm PMF}(k)}{2\pi^2} \nonumber\\   
&\approx 10^{-4}\,\xi_{\rm b}^2(a) \left(\frac{k}{\rm Mpc^{-1}}\right)^{2n_{\rm B}+10} \left(\frac{B_{\lambda}}{\rm nG}\right)^4   \nonumber\\
&\quad \times G_{n_{\rm B}}\, e^{-2k^2\lambda_{\rm D}^2},
\label{baryon pk}
\end{align}
where $G_{n_{\rm B}}$ is a dimensionless factor that depends on the spectral index $n_{\rm B}$.

By combining Eq.~\eqref{baryon pk} and Eq.~\eqref{lambda_D}, we obtain the amplitude of $\Delta_{\rm b}^{\rm PMF}(k)$:
\begin{equation}
\Delta_{\rm b}^{\rm PMF}(k = 1/\lambda_{\rm D}) = \xi_{\rm b}^2(a)\,e^{-2}\,\kappa_{n_{\rm B}}^{-4}\,G_{n_{\rm B}}.
\label{eq:delta_b_pmf}
\end{equation}

Eq.~\eqref{eq:delta_b_pmf} implies that while the amplitude of the induced baryon perturbations $\Delta_{\rm b}^{\rm PMF}$ is independent of the magnetic field strength, the uncertainty in $\kappa_{n_{\rm B}}$ translates directly into a significant uncertainty in this amplitude. To date, a definitive consensus on the precise value of $\kappa_{n_{\rm B}}$ has yet to be reached, necessitating calibration through high-resolution MHD simulations\footnote{While \citet{Ralegankar:2024arh} reported the value of $\kappa_{n_{\rm B}}$, their analysis was restricted to the specific case of $n_{\rm B} = -2$. Since the general dependence of this coefficient on the spectral index has not yet been established, we treat $\kappa_{n_{\rm B}}$ as an uncertain parameter in our analysis.}. In this study, we adopt the methodology of \cite{Ralegankar:2024ekl} introducing an order-one parameter $\eta$. This parameter is defined as the amplitude of the baryon power spectrum at the damping scale $\lambda_{\rm D}$ and a reference scale factor $a = 0.01$:
\begin{equation}
\Delta_{\rm b}^{\rm PMF}(k = \lambda_{\rm D}^{-1},\, a = 0.01) = \eta.
\end{equation}
Consequently, $\kappa_{n_{\rm B}}$ can be expressed in terms of $\eta$:
\begin{equation}
\kappa_{n_{\rm B}} = \left(\frac{G_{n_{\rm B}}\,\xi_{\rm b}^2(0.01)\,e^{-2}}{\eta}\right)^{1/4}.
\end{equation}

Determining the exact value of $\eta$ falls beyond the scope of this work. While \citet{Ralegankar:2024ekl} have discussed conservative estimates such as $\eta = 0.1$ and $\eta = 0.3$, we fix our fiducial value at $\eta = 0.3$. This choice is phenomenologically motivated: adopting $\eta = 0.1$ yields an amplitude for the matter power spectrum that is too heavily suppressed, rendering the PMF signatures too weak to leave a distinct imprint on structure formation.

The PMF-induced dark matter power spectrum $P_{\rm DM}^{\rm PMF}(k)$ can be obtained simply by replacing $\xi_{\rm b}(a)$ with $\xi_{\rm DM}(a)$. Furthermore, because PMF-induced perturbations and the standard primordial curvature perturbations from inflation are independent stochastic processes, they are entirely uncorrelated. Therefore, their power spectra can be linearly superimposed, i.e.,
\begin{equation}
P^{\rm tot}(k) = P^{\rm \Lambda CDM}(k) + P^{\rm PMF}(k).
\end{equation}
$ P^{\rm \Lambda CDM}(k)$ can be computed using the CLASS code \cite{2011JCAP...07..034B}.

\section{Numerical simulations}\label{sec:sim}

\subsection{Simulation code}\label{subsec:code}

We performed our simulations using a modified \textsc{Gadget-3} code (an updated version of the publicly available \textsc{Gadget-2} code \cite{Springel:2005mi})
described in detail in \cite{Liao:2022niz}. Below, we summarize its key features and refer interested readers to \cite{Liao:2022niz} for the details.

The code computes gravity using a Tree-Particle-Mesh (TreePM) hybrid algorithm \cite{Xu:1994fk} and solves gas hydrodynamics using the \textsc{sphgal} smoothed particle hydrodynamics (SPH) implementation \cite{Hu:2014goh}. Our SPH scheme employs the Wendland $C^4$ kernel with $N_{\rm ngb} = 100$ neighbors.

Radiative cooling, star formation and stellar feedback are modeled using the subgrid framework originally developed by Refs. \cite{Scannapieco:2005zf,Scannapieco:2006pv} and later improved by Ref. \cite{Aumer:2013gpa}. The model incorporates metal-dependent radiative cooling based on the tables from Ref. \cite{Wiersma:2008cs} and tracks the abundances of $11$ chemical elements (H, He, C, N, O, Ne, Mg, Si, S, Ca, and Fe) for all gas and star particle. Gas particles are stochastically converted into star particles if they are in a convergent flow ($\nabla \cdot \boldsymbol{v}_{\rm gas} \leq 0$), have a temperature $T \leq 12000~{\rm K}$, and exceed a density threshold of $\rho_{\rm gas} = 2.2 \times 10^{-24}~{\rm g}~{\rm cm}^{-3}$ (equivalent to a hydrogen number density $n_{\rm H} = 1~{\rm cm}^{-3}$). The subsequent stellar feedback includes energy and mass return from Type Ia and Type II supernova explosions and stellar winds from asymptotic giant branch (AGB) stars. The implementation of the supernova feedback model follows \citet{Nunez:2017}, and the adopted outflow velocity is set to $v_{\rm SN} = 4000~{\rm km}~{\rm s}^{-1}$.

We model supermassive black hole (SMBH) evolution using the `gadget' implementation described in \cite{Liao:2022niz}. In any friends-of-friends (FoF) halo \cite{Davis:1985rj} that reaches a mass of $M_{\rm FoF,~BH} = 10^{10}~h^{-1}{\rm M}_\odot$ and does not already contain an SMBH, the highest-density gas particle is converted into an SMBH particle with a seed mass of $M_{\rm seed,~BH} = 10^{5}~h^{-1}{\rm M}_\odot$. Once seeded, the SMBH grows via gas accretion, with the accretion rate calculated using the Bondi model \cite{Bondi:1952ni}, capped at the Eddington limit. A fraction of the accreted mass energy is radiated away, governed by a radiative efficiency of $\epsilon_{\rm r} = 0.1$. A subsequent fraction of this radiated energy, set by a feedback efficiency parameter $\epsilon_{\rm f} = 0.05$, is thermally coupled to neighboring gas particles \cite{Springel:2004kf}. To counteract the underestimation of dynamical friction caused by softened gravitational interactions, the SMBH is repositioned to the local potential minimum at each timestep. Two SMBHs merge if they are within each other's smoothing length and their relative velocity is below half the local sound speed. Since the dynamics of SMBH binaries are not the focus of this work, we did not employ the code's more advanced modules specifically designed for this purpose, such as non-softened gravity, post-Newtonian corrections, and circumbinary disk accretion models.

This code has been used for both idealized galaxy merger simulations \cite{Liao:2023zxx,Liao:2023jci} and cosmological simulations \cite{Mannerkoski:2021lal,Mannerkoski:2021hgr,Keitaanranta:2025ncl}, producing galaxies whose properties show good agreement with observations. Furthermore, in Appendix~\ref{ap:z0_obs_compare}, we demonstrate that our new simulations also reproduce key observed galaxy properties at $z=0$.

Our subgrid models and parameters were calibrated against low-redshift observations and have not been tuned for the high-redshift universe. The primary goal of this study is therefore not to perfectly reproduce all observations, but to perform a comparative analysis of how PMFs alter galaxy formation with respect to the standard $\Lambda$--Cold Dark Matter ($\Lambda$CDM) model. A systematic calibration and comparison against a broader range of observational data is deferred to future work.

\subsection{Simulation details}

\begin{table}[htbp]
\centering
\caption{The magnetic field strength ($B_{\lambda}$) and spectral index ($n_B$) parameters for the PMF models adopted in this study.}

\begin{tabular*}{0.8\columnwidth}{@{\extracolsep{\fill}}ccc}
\hline
Name & $B_{\lambda}$ [nG] & $n_B$ \\
\hline \hline
no PMF & - & - \\
B1nB29 & 1 & -2.9 \\
B05nB29 & 0.5 & -2.9 \\
B1nB2 & 1 & -2 \\
B05nB2 & 0.5 & -2 \\
B02nB2 & 0.2 & -2 \\
\hline
\end{tabular*}
\label{tab:sim_params}
\end{table}

\begin{figure}[htbp!]
\centering
\includegraphics[width=0.95\columnwidth]{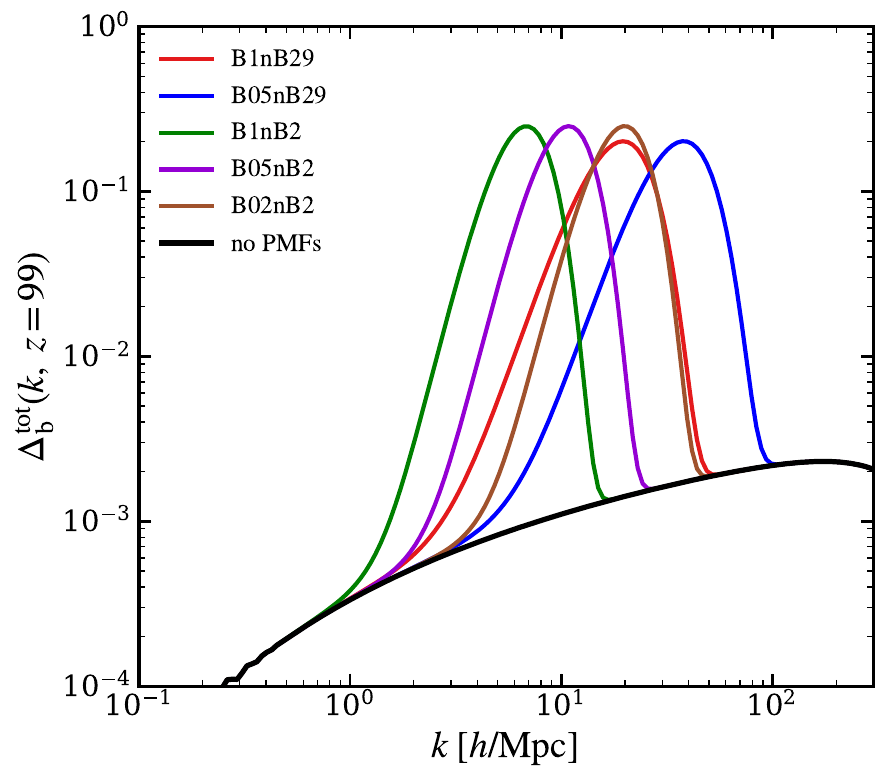}
\includegraphics[width=0.95\columnwidth]{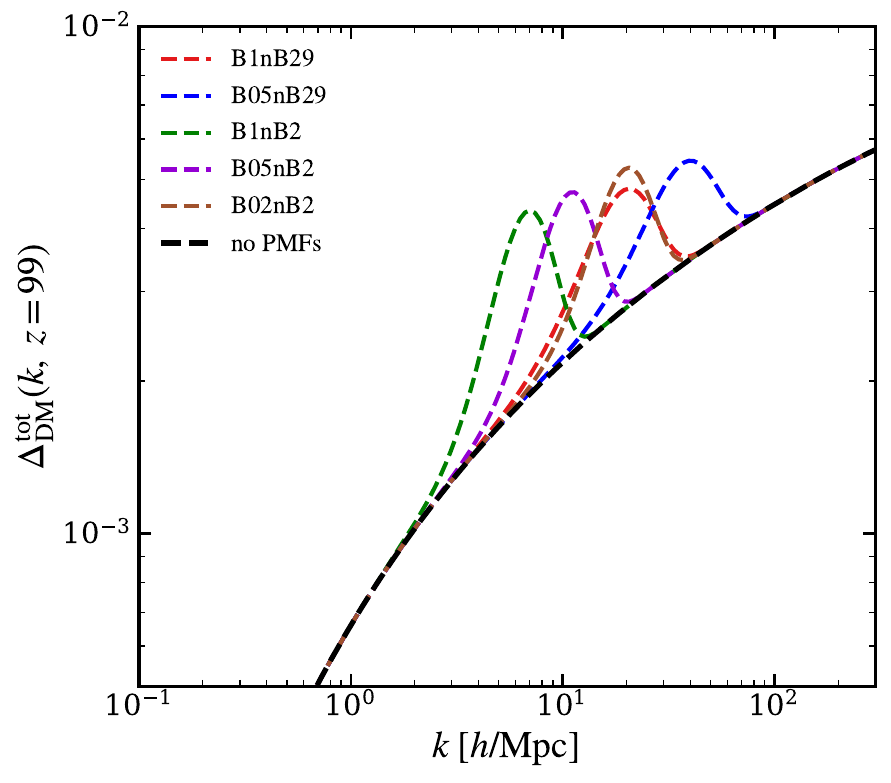}
\caption{Impact of PMFs on the dimensionless matter power spectra. The top and bottom panels show the baryon and dark matter power spectra at $z_{\rm IC} = 99$, respectively. In both panels, black lines denote the spectra without PMFs, while colored lines illustrate the effects of different PMF models.}
\label{fig:pk_b_dm}
\end{figure}

To investigate the impact of PMFs on structure formation, we performed six cosmological simulations: a standard $\Lambda$CDM model without PMFs (i.e., no PMF) and five models incorporating PMFs with parameters chosen to systematically probe the effects of magnetic field strength, spectral index, and peak scale (Table~\ref{tab:sim_params}). Specifically, four PMF models are from the combination of two magnetic field strengths ($B_\lambda = 0.5$, $1.0$ nG) and two spectral indices ($n_B = -2.9$ and $-2$). Their initial dimensionless power spectra,  shown in Fig.~\ref{fig:pk_b_dm}, have peak scales ($k_{\rm peak}$) from ${\sim} 6$ to ${\sim} 40~h{\rm Mpc}^{-1}$. As the impact of PMFs is sensitive to the peak scale \cite{Pavicevic:2025gqi}, we included an additional model ($B_\lambda = 0.2~{\rm nG}$, $n_{B} = -2$) specifically designed to have a similar peak scale ($k_{\rm peak} \sim 20~h{\rm Mpc}^{-1}$) as our $B_\lambda = 1~{\rm nG}$, $n_{B} = -2.9$ model. We refer to our PMF simulations using the convention `B[$B_{\lambda}$]nB[$|n_B|$]' where the bracketed values denote the magnetic field strength and the absolute value of the spectral index, omitting decimal points. While some of these PMF parameters may be in tension with current observational constraints (e.g., \cite{Pavicevic:2025gqi}), this pilot study intentionally includes models with stronger magnetic fields to maximize their impact on structure formation, thereby providing a clearer understanding of the underlying physical effects. A detailed comparison with observational constraints is reserved for future work.

Our simulations adopt cosmological parameters from the Planck 2015 results \cite{Planck:2015fie}, i.e., $\Omega_{\rm m} = 0.308$, $\Omega_{\rm b} = 0.04841$, $\Omega_\Lambda = 0.692$, $H_0 = 67.8~{\rm km}~{\rm s}^{-1}{\rm Mpc}^{-1}$ ($h = 0.678$), and $\sigma_8 = 0.8149$. Each simulation was initialized with $384^3$ dark matter particles and an equal number of gas particles within a periodic box of comoving side length $L_{\rm box} = 17~h^{-1}{\rm Mpc} \approx 25~{\rm Mpc}$. The particle number and box size were carefully chosen to resolve the scales where PMFs affect the initial matter power spectrum while maintaining a manageable computational cost. The resulting dark matter and gas particle masses are $m_{\rm DM} = 6.25 \times 10^{6}~h^{-1}{\rm M}_{\odot}$ and $m_{\rm gas} = 1.17 \times 10^{6}~h^{-1}{\rm M}_{\odot}$, respectively. We set the comoving gravitational softening length to $\epsilon_{\rm soft} = 0.88~h^{-1}{\rm kpc}$, corresponding to $1/50$ of the mean particle separation, following Ref. \cite{Zhang:2018nqh}. 

We generate the ICs for our simulations at $z_{\rm IC} = 99$ using a modified version of the \textsc{Ngenic} code \cite{Springel:2005nw,Crocce:2006ve,Angulo:2012ep}. As shown in Fig.~\ref{fig:pk_b_dm}, PMFs have a species-dependent influence on the initial power spectra, affecting baryons more significantly than dark matter, which is consistent with our theoretical discussion in Section \ref{sec:theory}. To properly incorporate this discrepancy, the initial phase space distributions for dark matter and gas must be generated separately. This precludes the use of the traditional method, where a single set of total matter particles is perturbed and subsequently split into dark matter and gas components (see \cite{Liao:2016lih} for an example). Instead, we adopt the method of \cite{Liao:2025bie} to generate a two-component particle load with equal number of particles for each species. The resulting distribution for each component, as well as for the total particle set, is a glass-like configuration that ensures homogeneity, isotropy, and minimal numerical noise.

We displaced the initial glass-like particle distribution using the Zel'dovich approximation to generate the particle positions and velocities. To ensure that the large-scale structures for dark matter and gas are spatially correlated, we used the same random seed to generate the initial displacement fields for both components. The perturbed particle positions ($\boldsymbol{x}$) are given by
\begin{equation}
    \boldsymbol{x} = \boldsymbol{q} - \nabla_{\boldsymbol{q}} \Phi(\boldsymbol{q}),
\end{equation}
where $\boldsymbol{q}$ is the initial Lagrangian (unperturbed) position and $\nabla_{\boldsymbol{q}} \Phi(\boldsymbol{q})$ is the displacement field. The corresponding peculiar velocities ($\boldsymbol{v}$) are
\begin{equation}
    \boldsymbol{v} = \frac{{\rm d}\boldsymbol{x}}{{\rm d}t} = - a H \frac{{\rm d}}{{\rm d}a} \nabla_{\boldsymbol{q}} \Phi(\boldsymbol{q}).
\end{equation}
In the standard approach, the time derivative on the right-hand side is typically calculated using a scale-independent growth factor. However, this is not applicable here, as PMFs induce a scale-dependent growth for which no general fitting formula exists. We therefore evaluate the derivative numerically by generating the displacement field at two slightly different redshifts and computing the finite difference. Specifically, we approximate the derivative as
\begin{equation}
    \frac{{\rm d}}{{\rm d}a} \nabla_{\boldsymbol{q}} \Phi(\boldsymbol{q}) \approx \frac{\nabla_{\boldsymbol{q}} \Phi(\boldsymbol{q}, z_{\rm IC}) - \nabla_{\boldsymbol{q}} \Phi(\boldsymbol{q}, z_{\rm IC}-\Delta z)}{a(z_{\rm IC}) - a(z_{\rm IC}-\Delta z)}.
\end{equation}
In this work, we adopted $\Delta z = 1$, having confirmed that our results are numerically converged with respect to this choice.

We evolved each simulation from its ICs down to a final redshift of $z = 0$. Throughout the evolution, we saved $21$ snapshots for post-processing analysis. Halos are identified in these snapshots using a standard FoF algorithm \cite{Davis:1985rj} with a linking length of $b = 0.2$ times the mean inter-particle separation. The algorithm returns all FoF groups containing at least 32 particles of any types. The center of each halo is defined as the position of the member particle with the minimum gravitational potential energy. We then define the virial radius, $R_{200}$, as the radius within which the mean matter density is $200$ times the cosmic critical density. The total mass within $R_{200}$ is the halo's virial mass, $M_{200}$. Finally, for each halo, we define a corresponding galaxy radius as $r_{\rm gal} = 0.1 R_{200}$. Unless otherwise specified, the galaxy properties are computed using the particles within $r_{\rm gal}$. 

Note that in this study, the primary focus is to investigate how initial PMF-induced density perturbations influence the subsequent formation and evolution of dark matter halos and galaxies. Rather than modeling full MHD dynamics throughout the whole simulation, magnetic field effects are incorporated solely via the initial power spectra. Crucially, our simulations are initialized after the epoch at which gravitational forces dominate over the Lorentz force induced by PMFs. Therefore, although the simulations do not track the subsequent MHD evolution of magnetic fields, this approximation is not expected to significantly affect the formation and evolution of dark matter halos. This methodology aligns with the framework established in \cite{Sanati:2020oay, Sanati:2024ijt, Ralegankar:2024ekl, Pavicevic:2025gqi}. By contrast, cosmological MHD simulations dedicated to the dynamical evolution of PMFs focus primarily on field-centric processes and properties, such as magnetic amplification or decay, spectral evolution, coherence scale, topology, helicity, and observational signatures like Faraday rotation \cite{Vazza:2014jga,Vazza:2017mbz,Vazza:2020phq, Mtchedlidze:2021bfy, Mtchedlidze:2022ewp,Mtchedlidze:2025fen,Schober:2026cyf}.

\section{Results}\label{sec:res}

\subsection{Visualization of matter distribution}

\begin{figure*}[htbp] 
    \centering
    \includegraphics[width=0.8\textwidth]{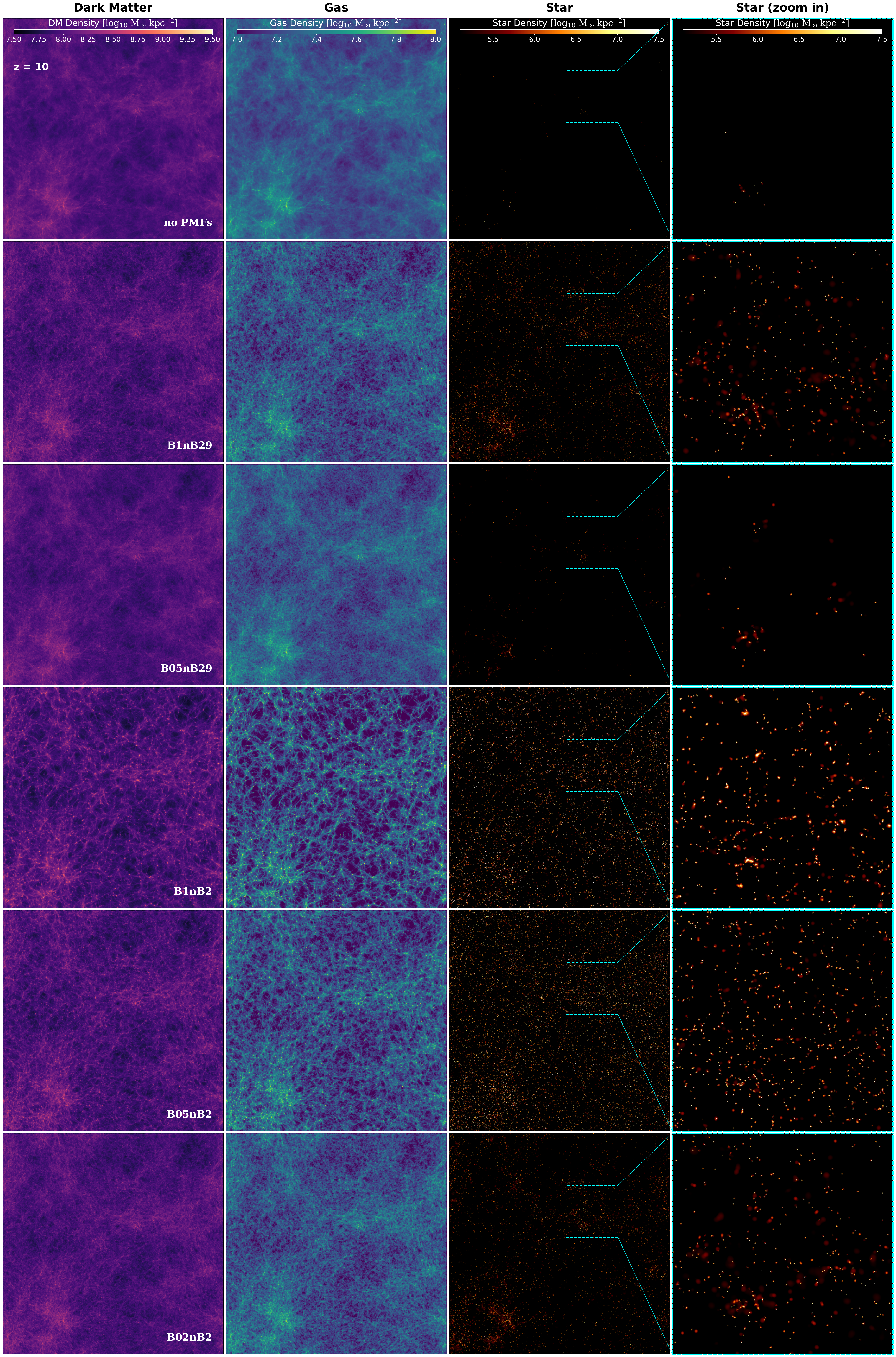}
    \caption{Matter distributions at $z=10$. The columns, from left to right, show the projected physical densities of the dark matter, gas, and stellar components across the simulation box. To illustrate the stellar distributions at smaller scales more clearly, the rightmost column displays a zoomed-in view of the region marked in the third column. Each row corresponds to a different model, as labeled on the left. The B1nB2 and B05nB2 models form more prominent filamentary structures earlier, due to their enhancement of the initial power spectrum at larger scales.}
    \label{fig:simu-visual-z10}
\end{figure*}

\begin{figure*}[htbp] 
    \centering
    \includegraphics[width=0.8\textwidth]{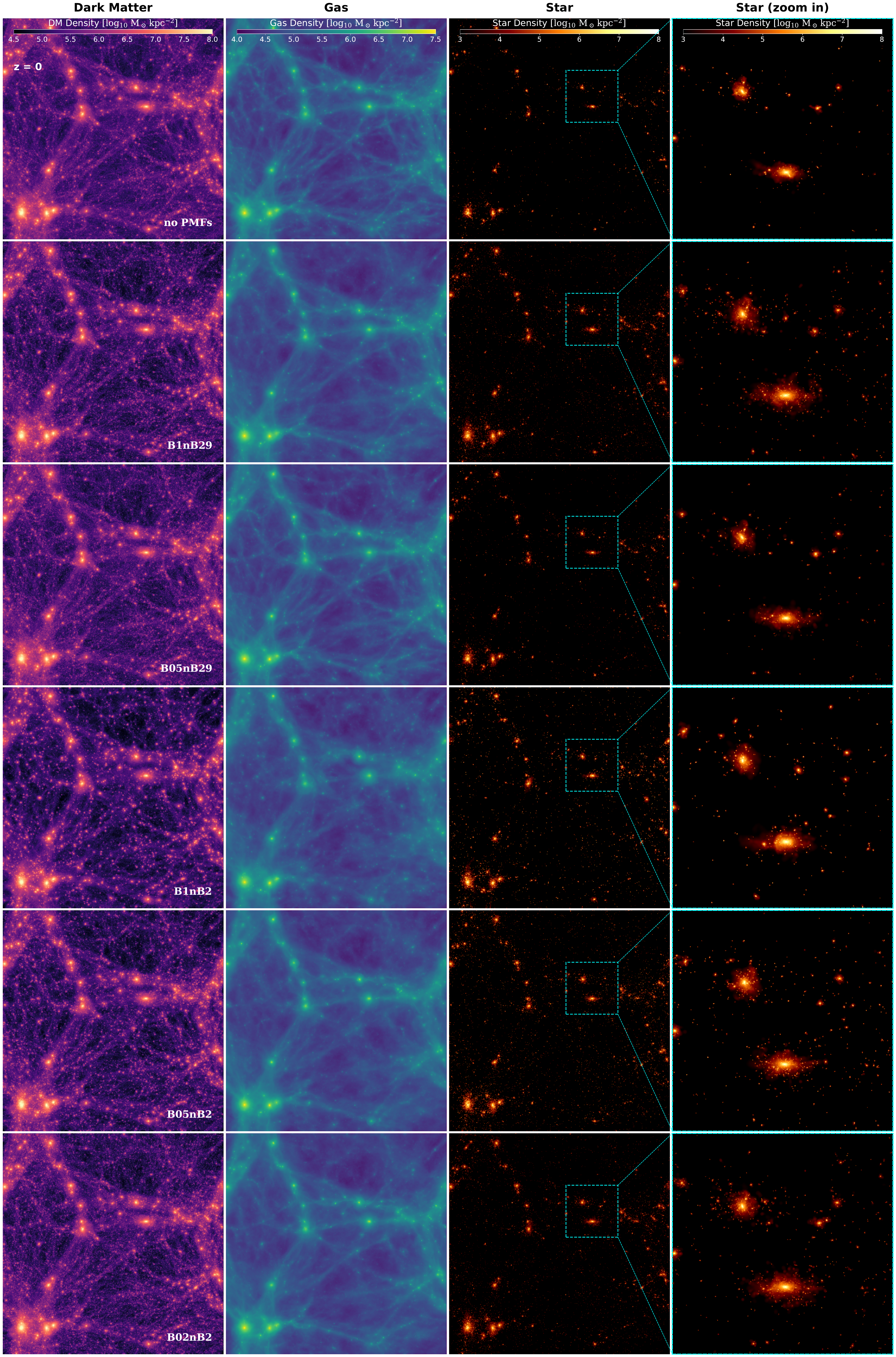}
    \caption{Similar as Fig.~\ref{fig:simu-visual-z10}, but showing the matter distributions at $z=0$. The overall differences between the models are less pronounced at $z=0$ than at $z=10$.}
    \label{fig:simu-visual-z0}
\end{figure*}

We start our analysis by looking at how PMFs influence the matter distribution within the simulation volumes. Figs. \ref{fig:simu-visual-z10} and \ref{fig:simu-visual-z0} present the two-dimensional projected surface density maps of dark matter, gas, and stars at high ($z=10$) and low ($z=0$) redshifts, respectively, for all simulation models listed in Table \ref{tab:sim_params}. The projection covers the entire simulation volume along the $z$-axis. 

At $z=10$, the impact of PMFs on structure formation is very significant. The inclusion of PMFs notably enhances the density of both dark matter and gas, leading to more pronounced structural features and an increased stellar population. This early enhancement of halo and star formation is in good agreement with recent findings by \citet{Ralegankar:2024ekl}, who observed a similar excess of halos and stars in their PMF simulations at $z=10$. This effect is particularly striking in the B1nB2 and B05nB2 models, which exhibit well-defined filamentary structures and voids. As indicated by the initial matter power spectra in Fig. \ref{fig:pk_b_dm}, these two models enhance perturbations at relatively larger scales compared to other PMF variants. Consequently, they facilitate the formation of more continuous and massive structures, with matter preferentially distributed along extended filaments.
In contrast, other PMF models featuring smaller magnetic field parameters (e.g., B05nB29) produce more isolated small-scale clumps. This morphological shift aligns perfectly with existing literature exploring weaker PMF regimes. For instance, \citet{Katz:2021iou} found that such PMFs lead to a strong enhancement of dark matter clumps at $z=6$, while rendering the filamentary structures significantly less smooth than in the $\Lambda$CDM case.
Furthermore, at this high-redshift stage, the influence of PMFs is more evident in the baryonic components (gas and stars) than in the dark matter. This is also consistent with the initial conditions shown in Fig. \ref{fig:pk_b_dm}, where the PMF-induced boost in the baryon power spectrum is significantly higher than that of the dark matter.

By $z=0$, the morphological differences introduced by PMFs become less prominent, though residual imprints remain discernible. Such persistent small-scale effects at $z=0$ were also reported by \citet{Ralegankar:2024ekl}, who observed an excess of bright spots in their PMF simulations. Similarly, \citet{Sanati:2020oay} found that  PMFs systematically increase the abundance of dark matter subhalos orbiting dwarf galaxies at this redshift. Interestingly, an opposite trend emerges compared to the high-redshift epoch: the discrepancies between the $\Lambda$CDM and PMF models are now more noticeable in the dark matter distribution than in the gas. This behavior is likely due to the dominance of baryonic feedback processes in the late universe (e.g., UV background heating, supernova feedback, and AGN feedback), which redistribute the gas and partially smooth out the initial density enhancements caused by PMFs. Since dark matter is collisionless and less sensitive to these processes (i.e., it only reacts due to the change of the gravitational potential), the structural differences imprinted at high redshifts are better preserved in the dark matter distribution. Additionally, the gas density in the PMF models at $z=0$ is slightly lower than in the $\Lambda$CDM case. This is because more efficient star formation in the PMF models, together with the resulting stronger feedback-driven redistribution, has reduced the gas remaining in dense regions.

\begin{figure*}[htbp!] 
    \centering
    \includegraphics[width=0.9\textwidth]{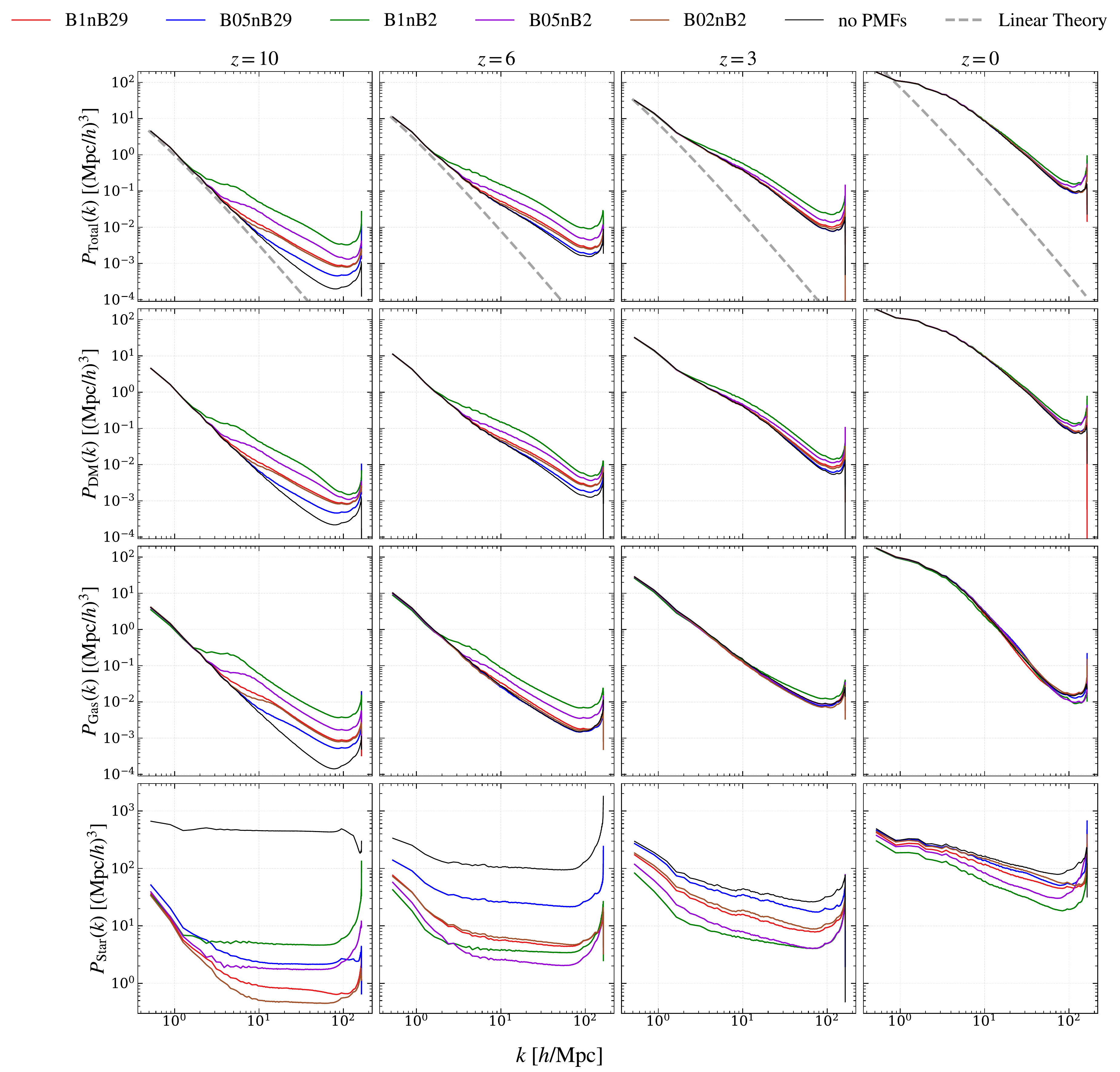}
    \caption{Matter power spectra for different PMF models computed using the package \texttt{Pylians} \cite{Pylians}. Each column shows a different redshift ($z=10$, $6$, $3$, and $0$, from left to right), and each row shows a different component (total, dark matter, gas, and stars, from top to bottom). The models are distinguished by color, as shown in the legend at the top. 
    In the first row (total matter), the gray dashed lines denote the linear power spectra for the $\Lambda$CDM model.
    For all components, the differences between PMF models are larger at higher redshifts.}
    \label{fig:pk_simulation}
\end{figure*}

\subsection{Matter power spectra}

To quantitatively characterize the PMF-induced modifications to the density field, Fig. \ref{fig:pk_simulation} displays the redshift evolution ($z=10,~6,~3,~0$)\footnote{As PMF effects are more prominent at high redshifts, in this study we present more comparison results at $z \geq 3$. See a similar approach in \cite{Ralegankar:2024ekl}.} of the auto-power spectra for total matter, dark matter, gas, and stars, comparing various PMF models with the baseline $\Lambda$CDM case. While the power spectra from our $L_{\rm box} = 25~{\rm Mpc}$ simulations are subject to cosmic variance on large scales, we mitigate this by focusing on the relative differences between models. Since all simulations share the same initial random seed and thus the same large-scale density phases, this approach cancels much of the variance and allows for a robust comparison of how PMFs influence structure formation. 

Across all components, the relative power spectrum enhancement induced by PMFs is most pronounced at high redshifts and systematically diminishes as the universe evolves towards $z=0$, consistent with the morphological evolution observed in the density maps. This trend reflects the transition from the linear regime to the late-time non-linear regime, where gravitational collapse and baryonic feedback processes increasingly diminish the initial imprints of the PMFs. Notably, the distinct peaks observed in the initial linear power spectra (as seen in Fig. \ref{fig:pk_b_dm}) are eventually redistributed into a broader enhancement in the nonlinear regime of the power spectrum, which arises from nonlinear mode coupling during hierarchical structure formation. Models that enhance the power spectrum on relatively larger scales in the ICs continue to leave discernible signatures on the power spectrum at those same scales even at low redshift. Furthermore, we find that models B1nB29 and B02nB2, which have similar $k_{\rm peak}$ in their initial dimensionless power spectra (see Fig. \ref{fig:pk_b_dm}), yield nearly identical results throughout the cosmic evolution.  
This suggests that the subsequent non-linear clustering is strongly governed by the characteristic scale of the initial seeds, leading to a significant degeneracy in how different PMF configurations impact the late-time matter distribution. This result echoes the finding in \cite{Pavicevic:2025gqi} that PMF models with a similar $k_{\rm peak}$ produce a similar increase in the 1D Ly$\alpha$ flux power spectrum.

For the dark matter and total matter power spectra, the scale-dependent signatures of PMFs remain consistent in their relative rankings across all epochs. Specifically, models that inject power at relatively larger scales in their ICs exhibit a stronger overall enhancement of the power spectrum across a broader range of scales throughout the subsequent non-linear evolution. However, the gas power spectrum in PMF models undergoes a late-time reversal, eventually falling slightly below the $\Lambda$CDM prediction because the enhanced star formation in these models leads to stronger cumulative feedback, which smooths the gas distribution at low redshifts.

Interestingly, the stellar power spectrum remains consistently lower in PMF models than in $\Lambda$CDM despite a higher total stellar mass. We interpret this as a result of clustering bias \cite{Mo:1995cs,Sheth:1999mn}. In the standard $\Lambda$CDM paradigm, star formation is highly biased, occurring predominantly within rare, massive halos at high-density peaks, leading to a highly clustered distribution and a high $P_{\text{star}}(k)$ amplitude. In contrast, the additional power from PMFs enables the ubiquitous collapse of lower-mass halos, allowing stars to form in a much more widespread and diffuse manner. This less-biased distribution results in a lower clustering amplitude.

\subsection{Abundance of halos and galaxies}

\begin{figure*}[htb!] 
    \centering
    \includegraphics[width=\textwidth]{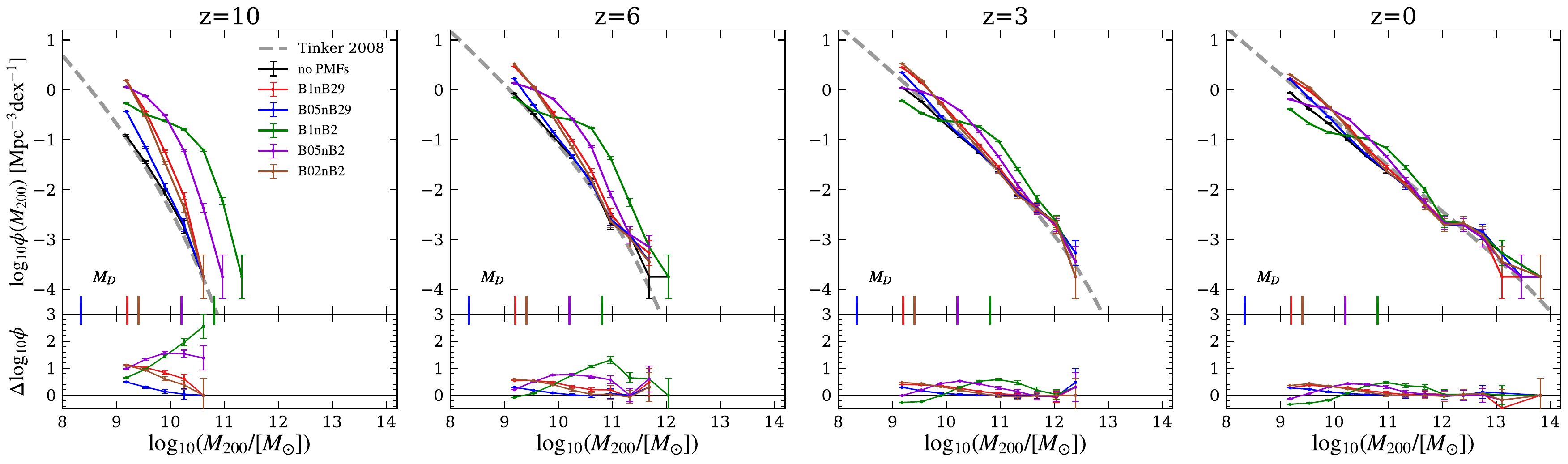}
    \caption{Halo mass functions at different redshifts ($z=10$, 6, 3, 0) for simulations with and without PMFs. Colored lines represent various PMF models and the black line shows the $\Lambda$CDM model. As a comparison, the gray dashed line shows the $\Lambda$CDM HMF computed using the fitting function of \citet{Tinker:2008ff}. The bottom sub-panels show the logarithmic difference relative to the $\Lambda$CDM case, i.e. $\Delta \log_{10}{\phi} \equiv \log_{10}{\phi}_{\rm PMF}(M_{200}) - \log_{10}{\phi}_{\Lambda{\rm CDM}}(M_{200})$, for each PMF model. Error bars show Poisson uncertainties. The vertical line segments mark the damping mass, $M_{\rm D}$, for different PMF models (see main text for details).}
    \label{fig:hmf}
\end{figure*}

\begin{figure*}[htb!] 
    \centering
    \includegraphics[width=\textwidth]{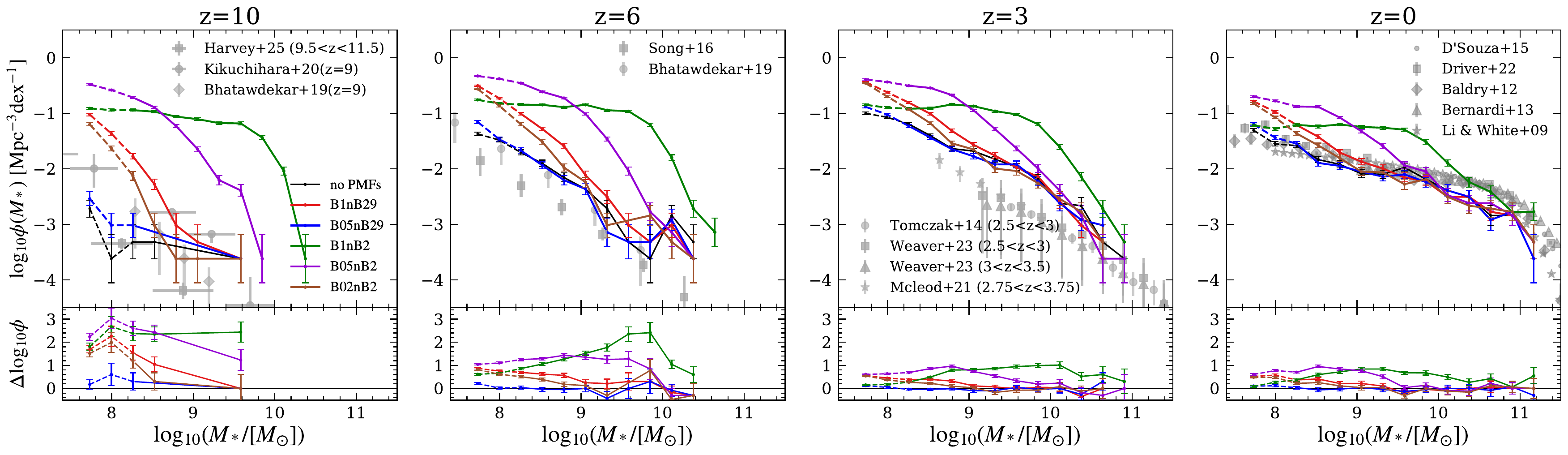}
    \caption{Stellar mass functions at $z=10$, $6$, $3$, and $0$ comparing various PMF models (colored lines) against the standard $\Lambda\mathrm{CDM}$ simulation (black line).  The bottom sub-panels show the logarithmic difference relative to the $\Lambda$CDM case, i.e. $\Delta \log_{10}{\phi} \equiv \log_{10}{\phi}_{\rm PMF}(M_{*}) - \log_{10}{\phi}_{\Lambda{\rm CDM}}(M_{*})$, for each PMF model.  Error bars show Poisson uncertainties. In this figure, we plot the SMF for all galaxies with at least 20 star particles, using dashed lines to mark the mass range below 100 star particles where numerical resolution effects may begin to play a role. Gray symbols indicate observational constraints: \citet{2015MNRAS.454.4027D}, \citet{Driver:2022vyh}, \citet{2012MNRAS.421..621B}, \citet{Bernardi:2013mqa}, and \citet{2009MNRAS.398.2177L} at $z=0$; \citet{2014ApJ...783...85T}, \citet{2023A&A...677A.184W}, and \citet{2021MNRAS.503.4413M} at $z=3$; \citet{bhatawdekar2019evolution} and \citet{song2016evolution} at $z=6$; and \citet{harvey2025epochs}, \citet{kikuchihara2020early}, and \citet{bhatawdekar2019evolution} at $z=10$. }
    \label{fig:smf}
\end{figure*}

In the following subsections, we further discuss the effects of PMFs on halo and galaxy properties. For our analysis of halo properties, we select all objects with $M_{200} \geq 10^{9}~M_\odot$, corresponding to a minimum of 100 dark matter particles. For the analysis of stellar properties, we focus on the galaxy sample with stellar mass $M_* \geq 2 \times 10^{8}~M_\odot$, which ensures at least 100 star particles per galaxy.

First, we examine their impact on the halo mass function (HMF) and stellar mass function (SMF) at different redshifts. Here, we consider the differential mass function, i.e., the abundance of halos or galaxies per unit logarithmic mass per unit comoving volume of the universe,
\begin{equation}
    \phi(M) \equiv \frac{1}{L_{\rm box}^3}\frac{{\rm d}N}{{\rm d}\ln M}.
\end{equation}

Fig. \ref{fig:hmf} shows the HMFs (at $z=10$, $6$, $3$, and $0$) for the different PMF models (colored lines), compared with the $\Lambda$CDM case (black line). The error bars represent the Poisson uncertainties derived from the number counts in each mass bin. In universes with stronger PMFs, massive halos form earlier. For example, at $z = 10$, the B1nB2 run has already produced halos with $M_{200} > 10^{11}~M_\odot$, which are absent in the $\Lambda$CDM run at that epoch.

As a more direct and quantitative comparison, the bottom subpanels of Fig. \ref{fig:hmf} plot the ratios of the HMFs from different PMF models to that of the $\Lambda$CDM model. Each PMF model enhances the HMF most strongly in the mass range corresponding to the scale where its initial power spectrum is boosted. To mark this scale, we use vertical line segments in Fig. \ref{fig:hmf} for the damping mass introduced by \citet{Ralegankar:2024ekl},
\begin{equation}
    M_{\rm D} \sim 7.4 \times 10^{12} \left(\frac{h/{\rm Mpc}}{k_{\rm D}}\right)^3 \frac{\Omega_{\rm m}h^2}{0.14} \left(\frac{0.678}{h}\right)^2 M_\odot h^{-1},
\end{equation}
where $k_{\rm D} \equiv 1 / \lambda_{\rm D}$. This damping mass roughly corresponds to the mass scale of the strongest HMF enhancement. 
The importance of this characteristic scale is also reflected in B1nB29 and B02nB2 models, which have similar $k_{\rm peak}$ and consequently show fairly similar HMF enhancement behaviors. Furthermore, the HMF enhancement is redshift-dependent, being stronger at high redshifts. For instance, the peak enhancement ratio in the B1nB2 model is $\sim 10^3$ at $z=10$, but decreases to $\sim 10^{0.5}$ by $z=0$. Toward lower redshifts, the early-collapse advantage produced by the PMF-enhanced initial power spectrum becomes less prominent, allowing the $\Lambda$CDM run to gradually catch up with the PMF runs. Therefore, the relative enhancement of HMFs in the PMF runs relative to $\Lambda$CDM run gradually decreases. At the low-mass end, this reduction of the relative excess can even turn into a mild deficit. At $z=0$, the HMFs for the B1nB2 and B05nB2 models are slightly lower than the $\Lambda$CDM prediction for masses $M_{200} \lesssim 10^{9.5}~M_\odot$. This mild suppression reflects the combined effect of the $\Lambda$CDM catch-up and the earlier assembly in the strong-PMF runs, where low-mass progenitors form earlier and are subsequently merged into more massive systems, leaving fewer residual low-mass halos at late times.

Fig. \ref{fig:smf} displays the SMFs from different PMF runs, which are also compared with a compilation of observational data from the literature (gray data points). Similar to HMFs, the enhancement in the SMF from PMFs decreases toward lower redshifts. However, the enhancement effect is more pronounced in the SMF, showing a stronger contrast between models especially at high redshifts. This is because baryonic processes, like cooling and star formation, amplify the underlying differences in the halo population.  
The role of this baryonic regulation is also evident from the comparison between the B1nB29 and B02nB2 models (with similar $k_{\rm peak}$).  Although these two models also produce comparable SMFs across different redshifts, their separation in the SMFs is slightly larger than in the corresponding HMFs.

The comparison with observational data reveals several key points. The $\Lambda$CDM run and the PMF runs with weaker fields (B1nB29, B05nB29, and B02nB2) are in broad agreement with observations across cosmic time. We note two points of minor tension: at $z=3$, our simulated SMFs are marginally higher than observed, and at $z=0$, our simulation box size leads to large uncertainties at the high-mass end, preventing us from clearly resolving the SMF `knee'. The PMF models with strong fields (B1nB2 and B05nB2) produce a distinct overabundance of galaxies at high redshifts, placing them in strong tension with these observations.

\subsection{Baryonic fractions and star formation}

\begin{figure*}[htb!] 
    \centering
    \includegraphics[width=\textwidth]{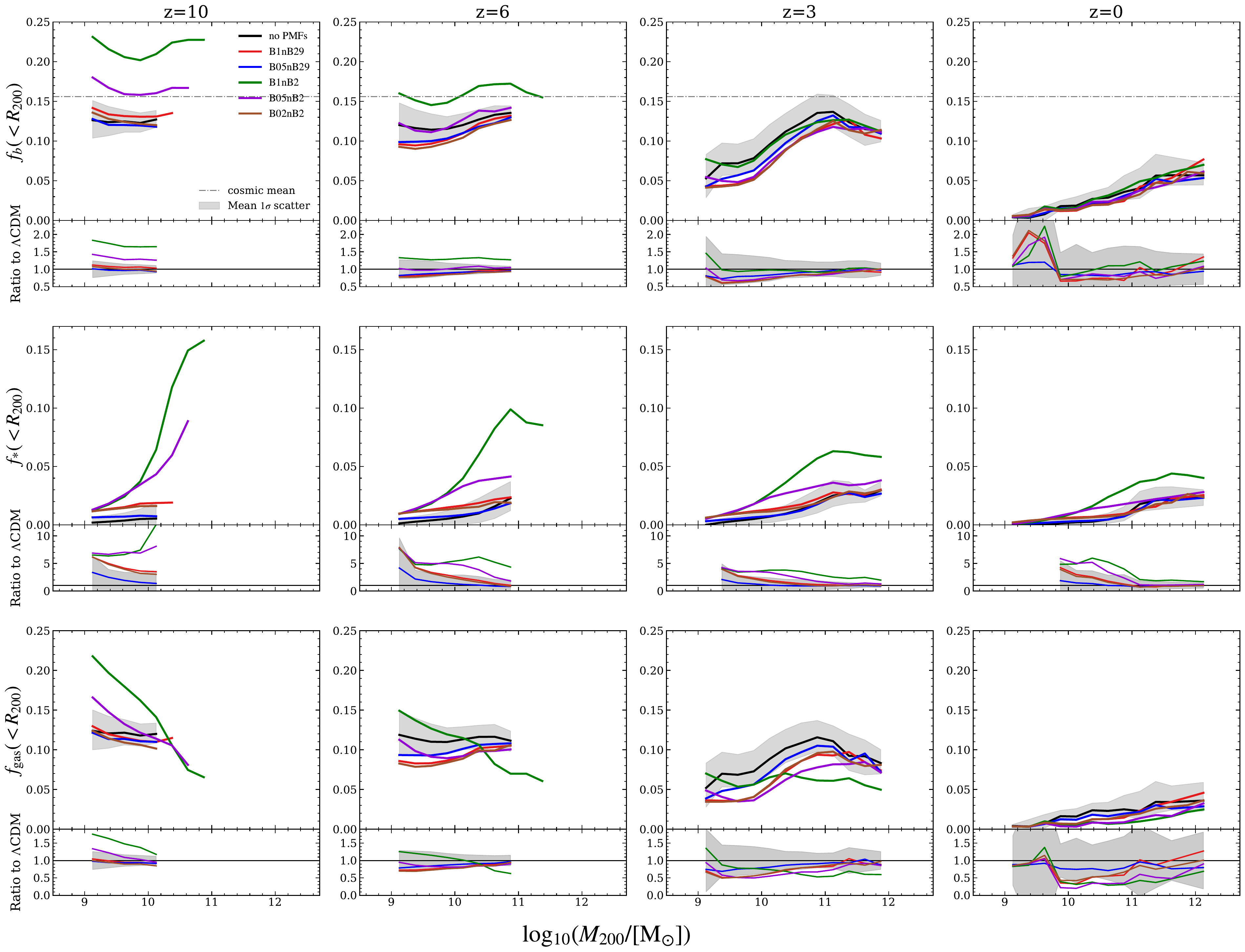}
    \caption{Redshift evolution of the baryon fraction ($f_{\rm b}$), stellar fraction ($f_*$), and gas fraction ($f_{\rm gas}$) within $R_{200}$ as a function of halo mass $M_{200}$. In each main panel, solid lines show the median relation for a given model, and the gray shaded regions represent the mean $1\sigma$ scatter (16th–84th percentile) centered on the $\Lambda$CDM baseline. All relations are shown only in mass bins containing at least 10 halos. The horizontal dash-dotted line in the $f_{\rm b}$ panels marks the cosmic mean baryon fraction ($\Omega_{\rm b} / \Omega_{\rm m}$). The sub-panels display the ratio of each PMF model to the $\Lambda$CDM result, where the gray shaded area indicates the propagated $1\sigma$ uncertainty of the ratio. To avoid division by zero, the ratios are computed only for mass bins with non-zero fractions in the $\Lambda$CDM simulation.}
    \label{fig:fraction}
\end{figure*}

\begin{figure*}[htb!] 
    \centering
    \includegraphics[width=\textwidth]{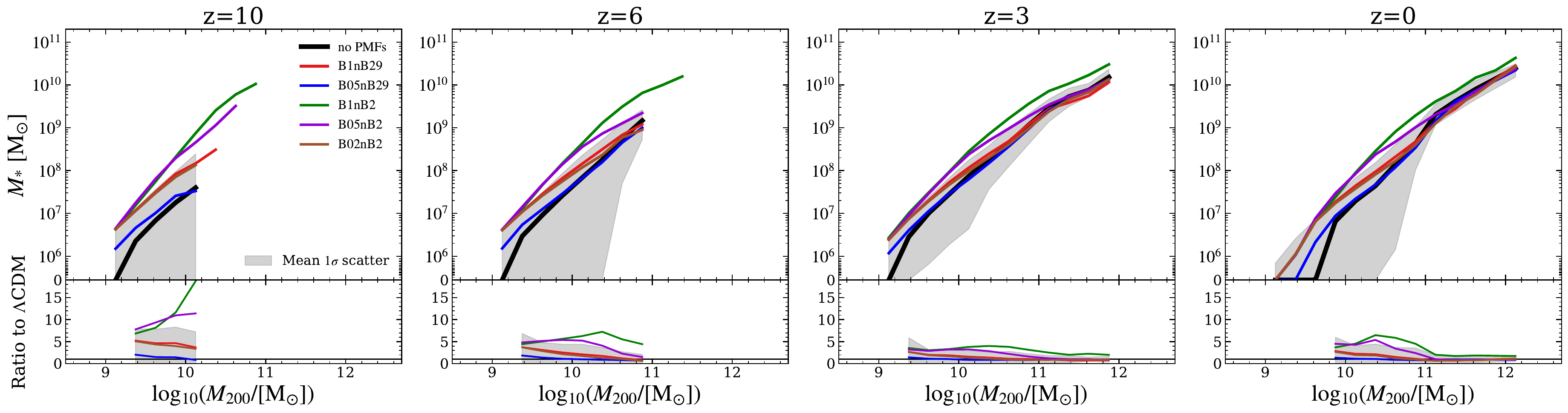}
    \caption{Redshift evolution of the stellar-to-halo mass ($M_* - M_{200}$) relation for the different PMF models at $z=10$, 6, 3, and 0. In each main panel, solid lines show the median relation for each model, and the gray shaded region denotes the mean $1\sigma$ scatter centered on the $\Lambda$CDM baseline. Note that for clarity, the $y$-axis uses a linear scale in the range of $0 \leq M_*/M_\odot \leq 10^{6}$ and a logarithmic scale for $M_* > 10^{6}~M_\odot$. All relations are shown only in mass bins containing at least 10 halos. The lower sub-panels show the ratio of the median $M_*$ in each PMF model to that in the $\Lambda$CDM run. Ratios are only computed where the corresponding $\Lambda$CDM model has a non-zero median stellar mass. The gray shaded area shows the propagated $1\sigma$ uncertainty of the ratio.    
    }
    \label{fig:star-halo}
\end{figure*}

\begin{figure*}[htb!] 
    \centering
    \includegraphics[width=\textwidth]{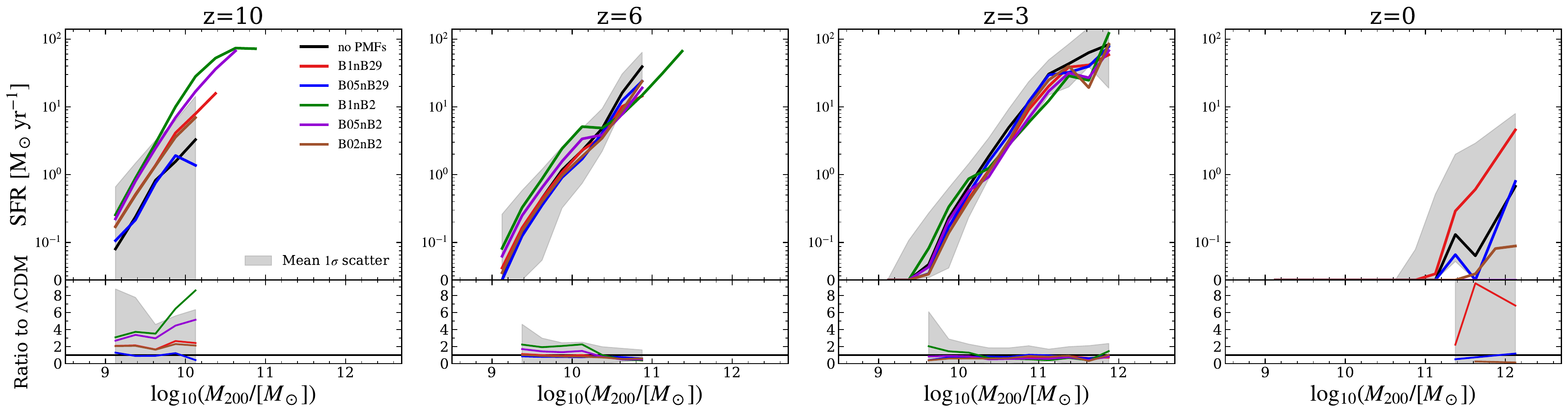}
    \caption{Redshift evolution of the star formation rate as a function of halo mass, ${\rm SFR}-M_{200}$, for different PMF models at $z=10$, 6, 3, and 0. The plotting layout, color scheme, and method for calculating the uncertainty (gray bands) are the same as in Fig.~\ref{fig:star-halo}.}
    \label{fig:SFR}
\end{figure*}

To further understand how PMFs influence galaxy formation, we examine the baryonic content in dark matter halos. Specifically, we examine the mass fractions of baryonic (`b'), gas (`gas'), and stellar (`$*$') components within the virial radius of each halo,
\begin{equation}
    f_X = M_X / M_{200},
\end{equation}
where $X = {\rm b}$, gas, or $*$, and $M_X$ is the total mass of the $X$ component within $R_{200}$.

The relation between these baryonic fractions and the halo mass is plotted in Fig. \ref{fig:fraction}. The solid lines show the median values for different models. The gray shaded regions indicate the mean $1\sigma$ scatter around the $\Lambda$CDM baseline. In the upper sub-panels, we first compute the median together with the 16th and 84th percentiles in each halo-mass bin for each model separately. The lower and upper deviations from the median are then measured for each model, averaged over all models, and finally recentered on the $\Lambda$CDM median to construct the shaded band. In the lower sub-panels, we show the ratio of each PMF model to the $\Lambda$CDM run. For each PMF model, the ratio in a given halo-mass bin is defined as $R = f_{\rm PMF} / f_{\Lambda{\rm CDM}}$, where $f_{\rm PMF}$ and $f_{\Lambda{\rm CDM}}$ are the corresponding median fractions of the PMF and $\Lambda$CDM models, respectively. Since propagating asymmetric uncertainties is not straightforward, we follow the procedure adopted in \cite{zheng2022impact}. Specifically, the asymmetric 16th$-$84th percentile range in each bin is first converted into an equivalent symmetric uncertainty following Method 2 in Appendix A of \cite{2017ChPhC..41c0001A}. The uncertainty of the ratio of each PMF model is then obtained through standard error propagation, $\sigma_{R} = |R| \sqrt{(\sigma_{\rm PMF} / f_{\rm PMF})^2 + (\sigma_{\Lambda{\rm CDM}} / f_{\Lambda{\rm CDM}})^2}$. Finally, the ratio uncertainties are averaged over all PMF models in each mass bin, and the resulting mean $1\sigma$ scatter is shown as a gray band centered on unity. The same plotting conventions and uncertainty estimations are adopted for subsequent figures (Figs. \ref{fig:star-halo}, \ref{fig:SFR}, \ref{fig:BH-M200} and \ref{fig:BH-M*}), and will not be repeated in the following discussion.

As shown in the top row, at $z=10$, halos in PMF models generally exhibit larger baryon fractions, $f_{\rm b}$, than those in the $\Lambda$CDM run. This enhancement is particularly pronounced in the B1nB2 and B05nB2 models. This trend is physically expected, because PMFs enhance the baryon density perturbations more strongly than the dark matter perturbations in the ICs (see Fig. \ref{fig:pk_b_dm}). This `baryon-biased' power enhancement allows baryons to collapse into halo potential wells more efficiently at early times. As redshift decreases, the differences in $f_{\rm b}$ become smaller, indicating that the PMF imprint on the total baryon content is progressively weakened by non-linear gravitational evolution and late-time astrophysical processes. At lower redshifts, the baryon fractions in almost all the PMF runs can be slightly lower than in $\Lambda$CDM, as a result of the lower gas fractions driven by stronger feedback effects from early star formation and black hole growth (see the detailed discussion of $f_{\rm gas}$ below). Compared with \cite{Ralegankar:2024ekl}, the enhancement of $f_{\rm b}$ in our simulations is more moderate. Specifically, \citet{Ralegankar:2024ekl} found that the B1nB29, B05nB29, and B02nB2 models exhibit baryon fractions above the cosmic mean for halos with masses $10^{9.5} \lesssim M_{\rm halo}/M_\odot\lesssim 10^{11}$ at $z = 10$ and $z = 4$. In our runs, however, the baryon fractions in these models are below the cosmic mean value across the whole plotted mass range. The reason is that our simulations incorporate stronger stellar feedback and include AGN feedback, both of which effectively reduce baryon retention within halos. This demonstrates that it is critical to consider the impact of baryonic subgrid models when quantifying PMF effects on structure formation.

The enhanced baryon supply and earlier halo collapse in the PMF models promote more efficient star formation, leading to higher stellar fractions ($f_*$), as shown in the middle row of Fig.~\ref{fig:fraction}. This enhancement is most pronounced at higher redshift and decreases over time. However, unlike the baryon fractions, the stellar fractions in all PMF models remain elevated compared to the $\Lambda$CDM run, even at $z=0$, and the differences are particularly significant in the stronger PMF models. This suggests that the stellar component retains a longer memory of the initial PMF-induced enhancement. This effect is also evident in the stellar-to-halo mass relation (SHMR), a standard diagnostic for galaxy formation efficiency. We present the SHMR for our models in Fig.~\ref{fig:star-halo}, showing a consistent enhancement for the PMF runs.

The evolution of the gas fraction is somewhat more complex, as shown in the bottom row of Fig.~\ref{fig:fraction}. At $z=10$, while most PMF models already show a reduced $f_{\rm gas}$ due to efficient star formation, the strong-PMF models, B1nB2 and B05nB2, remain notably above the $\Lambda$CDM relation. This is because that their profound initial enhancement in total baryon content is sufficient to overcompensate for the rapid gas consumption during the earliest stages of galaxy assembly. However, toward lower redshifts, even these high-PMF models eventually fall below the $\Lambda$CDM predictions and a clear trend emerges: models with a larger early-time enhancement of $f_{\rm b}$ tend to have lower $f_{\rm gas}$ at later times. This trend suggests more rapid baryon processing in the stronger PMF runs. The earlier collapse leads to earlier star formation and faster gas consumption. Simultaneously, the cumulative effect of enhanced stellar and AGN feedback, fueled by the early growth of stars and SMBHs, more effectively heats the interstellar medium and expels gas beyond $R_{200}$. This reduced gas mass directly drives the lower low-redshift baryon fractions in PMF models compared to $\Lambda$CDM, as seen in the top row. It is worth noting that, at low redshift, $f_{\rm gas}$ and $f_*$ provide a clearer separation among different PMF models than $f_{\rm b}$.

As a further diagnostic, Fig.~\ref{fig:SFR} shows the median star formation rate (SFR)--halo mass relations for different models. At high redshift ($z=10$), the larger gas reservoirs in the stronger PMF models fuel SFRs several times higher than in the $\Lambda$CDM counterpart. As the universe evolves, however, stronger baryonic feedback in the PMF models causes the SFRs in massive halos to drop, eventually falling below the $\Lambda$CDM predictions. By $z=0$, this effect is so pronounced that most galaxies in the strong-PMF models (B1nB2 and B05nB2) have already quenched, leaving them with zero median SFRs.  In contrast, the weaker PMF and $\Lambda$CDM models still show active star formation in massive halos. This evolution of the SFR--halo mass relation supports the picture of the PMF impact on baryonic fractions outlined above.

\subsection{Supermassive black holes}
 
\begin{figure*}[htb!] 
    \centering
    \includegraphics[width=\textwidth]{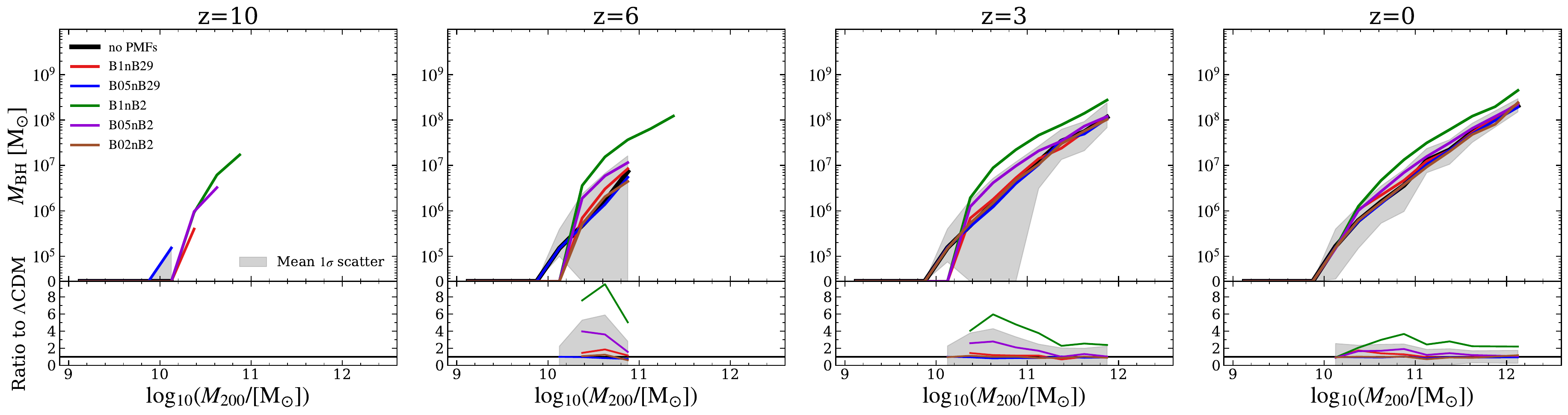}
    \caption{Redshift evolution of the SMBH-halo mass relation, $M_{\rm BH}-M_{200}$, for the different models. We adopt the same plotting layout, color scheme, and method for calculating the uncertainty (gray bands) as in Fig.~\ref{fig:star-halo}.}
    \label{fig:BH-M200}
\end{figure*}

\begin{figure*}[htb!]
    \centering    
    \includegraphics[width=\textwidth]{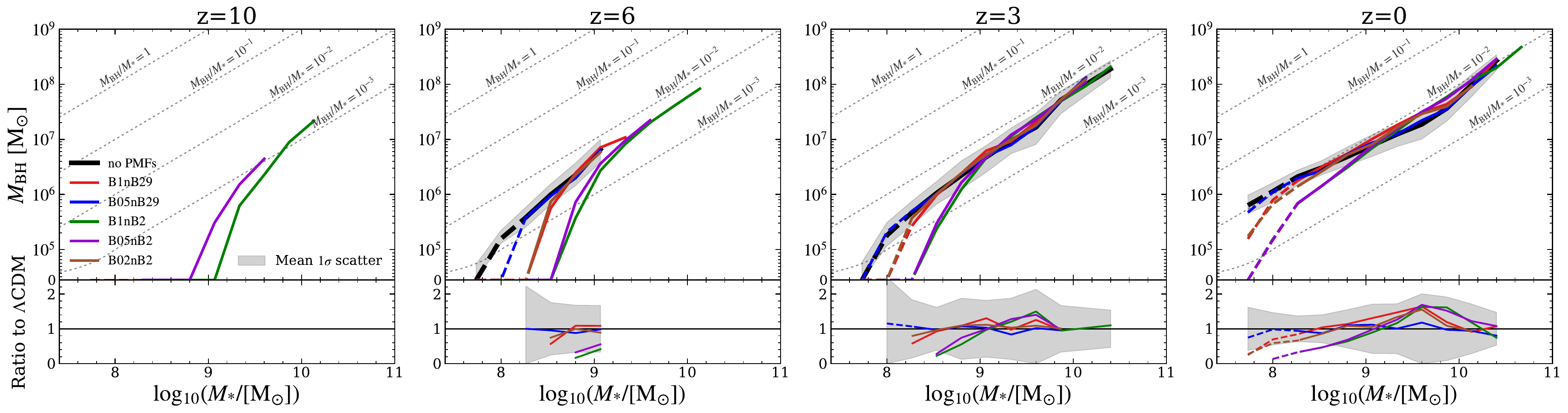}
    \caption{Evolution of the SMBH-stellar mass relation, $M_{\rm BH}-M_*$, for  the different models at $z=10$, 6, 3, and 0. In each panel, solid lines show the median relation, with dashed portions indicating a resolution below 100 star particles (as in Fig.~\ref{fig:smf}). The overall layout, color scheme, and shaded uncertainty regions follow the conventions of Fig.~\ref{fig:star-halo}. For reference, the gray dotted lines represent constant SMBH-to-stellar mass ratios of $M_{\rm BH}/M_* = 10^{-3}$, $10^{-2}$, $10^{-1}$, and $1$.}
    \label{fig:BH-M*}
\end{figure*}

Unlike previous galaxy formation simulations with PMFs, our work incorporates SMBH accretion and feedback. We now leverage this advancement to examine the evolution of SMBHs, comparing the $M_{\rm BH}-M_{200}$ and $M_{\rm BH}-M_*$ relations across our simulations in Figs.~\ref{fig:BH-M200} and \ref{fig:BH-M*}, respectively.

As shown in the left panel of Fig.~\ref{fig:BH-M200}, at $z=10$, the strong-PMF models (B1nB2 and B05nB2) already host SMBHs with masses $M_{\rm BH} > 10^6~M_\odot$.  In contrast, the SMBHs in other runs have not yet grown to significant masses. This indicates that strong PMFs facilitate the rapid growth of SMBHs in the early universe. This finding is reminiscent of the observations of high-redshift luminous quasars (see e.g., \cite{Inayoshi:2019fun,Fan:2023araa}, for reviews) and the recent James Webb Space Telescope (JWST) detections of broad-line AGNs (the so-called `little red dots') at $z \gtrsim 5$ (e.g., \cite{Harikane:2023apj,Kocevski:2023apjl,Maiolino:2023bpi,Matthee:2023utn}), which suggest the early universe hosted surprisingly massive black holes. However, we must caution that our simulations do not resolve the initial black hole formation process. Instead, SMBHs are seeded with a fixed mass of $10^{5}h^{-1}M_\odot$ once their host FoF group reaches a mass of $10^{10}h^{-1}M_\odot$. Therefore, the presence of more massive SMBHs in our PMF runs is not a statement on SMBH seeding mechanisms, but rather a direct consequence of earlier structure formation and a more abundant gas supply available to feed the central black holes. The $M_{\rm BH}-M_{200}$ relations at lower redshifts are plotted in the remaining panels. At a given halo mass, SMBHs in stronger PMF models remain more massive as the universe evolves, although their mass enhancement relative to the $\Lambda$CDM model decreases over time.

Fig.~\ref{fig:BH-M*} displays the median relation between SMBH mass and host galaxy stellar mass.
For SMBHs with $M_{\rm BH} \lesssim 10^{6}~M_\odot$, the host stellar mass at fixed $M_{\rm BH}$ depends sensitively on the PMF strength, with stronger PMFs leading to more massive host galaxies. In contrast, for SMBHs with $M_{\rm BH} \gtrsim 10^{6}~M_\odot$, almost all models converge onto the same relation. This behavior occurs because when SMBHs are close to their seed mass, the accretion rates are low (in the Bondi model, accretion is proportional to $M_{\rm BH}^2$) and consequently the AGN feedback (proportional to SMBH accretion rate) are weak. With AGN feedback being ineffective, stellar mass grows much more efficiently than the central SMBH and is mainly driven by the higher gas supply and star formation rates originating from PMF enhancement.
However, once an SMBH becomes more massive ($M_{\rm BH} \gtrsim 10^{6}~M_\odot$), AGN feedback is stronger and becomes effective at regulating star formation. This feedback enforces a tight co-evolutionary relationship between the SMBH and its host galaxy, causing all models to converge onto a similar $M_{\rm BH}-M_*$ relation. This convergence indicates that the physics of galaxy-SMBH co-evolution overwhelms the initial PMF effects at the high-mass end.

At high redshifts ($z \gtrsim 3$), galaxies in our simulations have SMBH-to-stellar mass ratios in the range of $10^{-3} \lesssim M_{\rm BH}/M_* \lesssim 10^{-2}$ (Fig.~\ref{fig:BH-M*}). This means our SMBHs are not `overmassive' in the way reported by recent JWST observations, where ratios often exceed $0.01$ (e.g., \cite{Maiolino:2023bpi,Pacucci:2023oci,Jones:2025arXiv251007376J}). This result is a known consequence of the specific seeding and AGN feedback models adopted, a feature common to many current cosmological simulations \cite{Habouzit:2021}.

\section{Conclusion and Discussion}\label{sec:summary}

Primordial magnetic fields (PMFs) can alter structure formation by introducing a scale-dependent enhancement to the post-recombination matter power spectrum. This enhancement is stronger for baryons than for dark matter, because baryons are directly affected by the Lorentz force whereas dark matter responds only indirectly through gravity. In this work, we investigated how this PMF-induced excess power affects structure formation by performing a suite of cosmological hydrodynamical simulations that combine PMF-enhanced initial conditions with more comprehensive baryonic feedback models. 

Our main findings can be summarized as follows:

(i) {\it Matter distributions.} PMF models with with peak enhancement on larger scales (i.e., smaller $k_{\rm peak}$) produce more pronounced density structures, richer stellar components, and stronger enhancement in the total matter and dark matter power spectra. In particular, strong-PMF models (B1nB2 and B05nB2) form more continuous and massive filamentary structures, contrasting with the more isolated small-scale clumps seen in other models (Figs.~\ref{fig:simu-visual-z10} and \ref{fig:simu-visual-z0}).

(ii) {\it Evolution of matter power spectra.} Over time, the distinct initial peaks in the matter power spectrum are redistributed by nonlinear evolution into a broad excess of power. Models with peak enhancement at larger scales (smaller $k_{\rm peak}$) preserve stronger signatures over a wider range of scales. In addition, the responses of different components are not identical. At $z=10$, the PMF imprint is more evident in the baryonic components, whereas by $z=0$, the residual differences are more noticeable in the dark matter distribution than in the gas, because baryonic feedback redistributes and smooths the gas more efficiently. Moreover, unlike the power spectra of the other components, the stellar power spectra in PMF runs remain lower than in the $\Lambda$CDM case despite the richer stellar component. This suggests that PMFs make star formation less biased and more spatially extended (Fig.~\ref{fig:pk_simulation}).

(iii) {\it Halo and galaxy abundances.} PMFs can shift halo growth, baryon collapse, and star formation to earlier times. This leads to a significant enhancement in both the HMF and SMF at high redshift, particularly at the mass scales corresponding to the scales where the initial power spectra are boosted. The effect is more pronounced in the SMF, indicating that baryonic processes like cooling and star formation amplify the initial PMF-driven enhancement in the halo population (Figs.~\ref{fig:hmf} and \ref{fig:smf}). 

(iv) {\it Baryonic fractions.} Consistently, at high redshift, the PMF runs also have higher baryon and stellar fractions, together with larger stellar masses and higher SFRs at fixed halo mass. This follows from the stronger PMF enhancement in baryons, which allows gas to collapse into halo potential wells more efficiently and enables star formation in lower-mass halos that remain star-free in the $\Lambda$CDM case. At later times, however, the stronger-PMF models develop lower gas fractions and much weaker SFR enhancement, showing that the early boost is reprocessed by faster gas consumption and stronger cumulative stellar and AGN feedback (Figs.~\ref{fig:fraction}, \ref{fig:star-halo}, and \ref{fig:SFR}). 

(v) {\it Impact on SMBH growth.} Strong-PMF models produce SMBHs in earlier time and show larger $M_{\rm BH}$ at fixed halo mass. For the SMBH-stellar mass relation, at low masses ($M_{\rm BH} \lesssim 10^6~M_\odot$), PMF enhancements boost the growth of stellar mass over the central SMBH due to initially weak AGN feedback. However, this PMF-driven difference is erased once SMBHs become massive enough for their powerful AGN feedback to enforce a tight co-evolutionary relationship, causing all models to converge onto the same $M_{\rm BH}$--$M_*$ relation (Figs.~\ref{fig:BH-M200} and \ref{fig:BH-M*}).

(vi) {\it Redshift dependence of PMF effects.} The differences between the PMF and $\Lambda$CDM runs are generally most pronounced at high redshift and gradually weaken toward low redshift because of non-linear gravitational evolution and baryonic feedback. Nevertheless, these imprints are not completely erased by late times, and some quantities still retain visible offsets, indicating that the impact of PMF-enhanced initial conditions can leave detectable signatures even at low redshifts.

(vii) {\it Effects of peak scales.} The impact of PMFs depends strongly on the characteristic peak scale, $k_{\rm peak}$, of the initial power enhancement. Models with similar $k_{\rm peak}$ can yield very similar non-linear evolution even when their PMF parameters differ. This is illustrated by the B1nB29 and B02nB2 models, which show nearly identical evolution in almost all of the quantities examined in this paper despite their different magnetic-field strengths and spectral indices.

Our results also suggest which observables are likely to be most useful for studying PMFs. At high redshift, nearly all of the quantities studied here show clear PMF-induced differences, indicating that the early Universe is the most sensitive regime for detecting PMF signatures. In this sense, high-redshift observational experiments, such as JWST \cite{2006SSRv..123..485G,2024ApJ...972..143C}, may provide a promising avenue for probing and constraining PMFs \cite{Zhang:2024yph,Fairbairn:2026dva}. At low redshift, stellar-related quantities may remain especially useful, because the PMF imprint on the stellar component is more persistent. In particular, the stellar power spectrum, SMF, stellar fraction and the $M_{*}-M_{200}$ relation still show significant offsets. 
Besides, $f_{\rm gas}$ is also informative because it shows a characteristic late-time suppression in the stronger-PMF models. Finally, SMBH related observables may provide an additional handle on stronger PMFs. 

Note that in our simulation initial condition, we consider only PMF-induced scalar (density) perturbations, omitting vector (vorticity) and tensor (gravitational waves) perturbations (see \cite{Subramanian:2015lua} for a review). Because the formation and growth of halos are primarily driven by scalar density perturbations and their non-linear gravitational evolution, the omission of the vector perturbations is not expected to significantly affect our primary conclusions regarding halo structure and abundance. Furthermore, \citet{Ralegankar:2024ekl} demonstrated that PMF-induced vortical motions have a negligible impact on the halo baryon fraction. While they noted that vortical motions can reduce star formation rates at very high redshifts ($z > 10$) by enhancing halo angular momentum support, this effect diminishes significantly at $z \lesssim 10$. Given that our study focuses on $z \leq 10$, the impact of omitting initial vector perturbations on our results regarding star formation and galaxy properties should be minor. We leave the inclusion of the full PMF-induced vortical velocity field to future work.

Although our simulations adopt more comprehensive baryonic physical models, it should be noted that we do not perform full MHD simulations. In our setup, PMFs are included only through their effect on the initial conditions. This approximation is expected to have small impact on quantities that are mainly governed by the PMF-modified density field and subsequent gravity, such as HMF, total and dark matter power spectra and high-redshift baryon fraction. By contrast, it may play a non-negligible role in shaping observables that depend sensitively on the later baryonic evolution, since magnetic fields can alter the gaseous response and thereby modify star formation and stellar buildup \cite{2018MNRAS.473.4077P, 2017MNRAS.467..179G, Katz:2021iou, Marinacci:2015ria, Sanati:2024ijt}. Meanwhile, cosmological MHD studies have shown that the magnetic field itself evolves during structure formation, with its strength, topology, power spectrum, and correlation length modified by gravitational collapse and MHD turbulence \cite{Vazza:2014jga,Vazza:2017mbz,Vazza:2020phq, Mtchedlidze:2021bfy, Mtchedlidze:2022ewp,Mtchedlidze:2025fen,Schober:2026cyf}. Such dynamical magnetic evolution may, in turn, affect the subsequent star formation and galaxy evolution. Additionally, investigating how magnetic field properties evolve within different cosmic web environments offers a powerful probe of their primordial origins (see e.g., \cite{Vazza:2020phq,Mtchedlidze:2021bfy,Mtchedlidze:2022ewp,Mtchedlidze:2024kvt,Hutschenreuter:2018vkr}). In future work, large-volume cosmological simulations combining our comprehensive galaxy formation subgrid models with a full MHD treatment will be essential to capture both sets of physical processes simultaneously.

\section*{Acknowledgements}
We thank the anonymous referee for helpful comments. We thank Daneng Yang, Pranjal Ralegankar, Ziwei Wang, Lei Lei, and Chi Zhang for their useful discussions. This work is supported by the National Key Research and Development Program of China (No. 2022YFF0503304) and the National Natural Science Foundation of China (Nos. 12588202, 12473015, and No. 12220101003). 

\section*{Data Availability}
The data that support the findings of this article are not publicly available. The data are available from the corresponding authors upon reasonable request.

\appendix

\section{Comparison with observations at $z=0$}\label{ap:z0_obs_compare}

The details of our baryonic subgrid models are described in Section~\ref{subsec:code}. The parameters for these models were calibrated within the $\Lambda$CDM framework to reproduce key galaxy properties at $z \sim 0$, and our specific implementation has been extensively validated in previous work \citep{Liao:2023zxx,Liao:2023jci,Mannerkoski:2021lal,Mannerkoski:2021hgr,Keitaanranta:2025ncl}. As a further validation, this appendix compares the $z=0$ results from the simulations performed in this work against key observations.

\begin{figure*}[htb!] 
    \centering
    \includegraphics[width=0.9\textwidth]{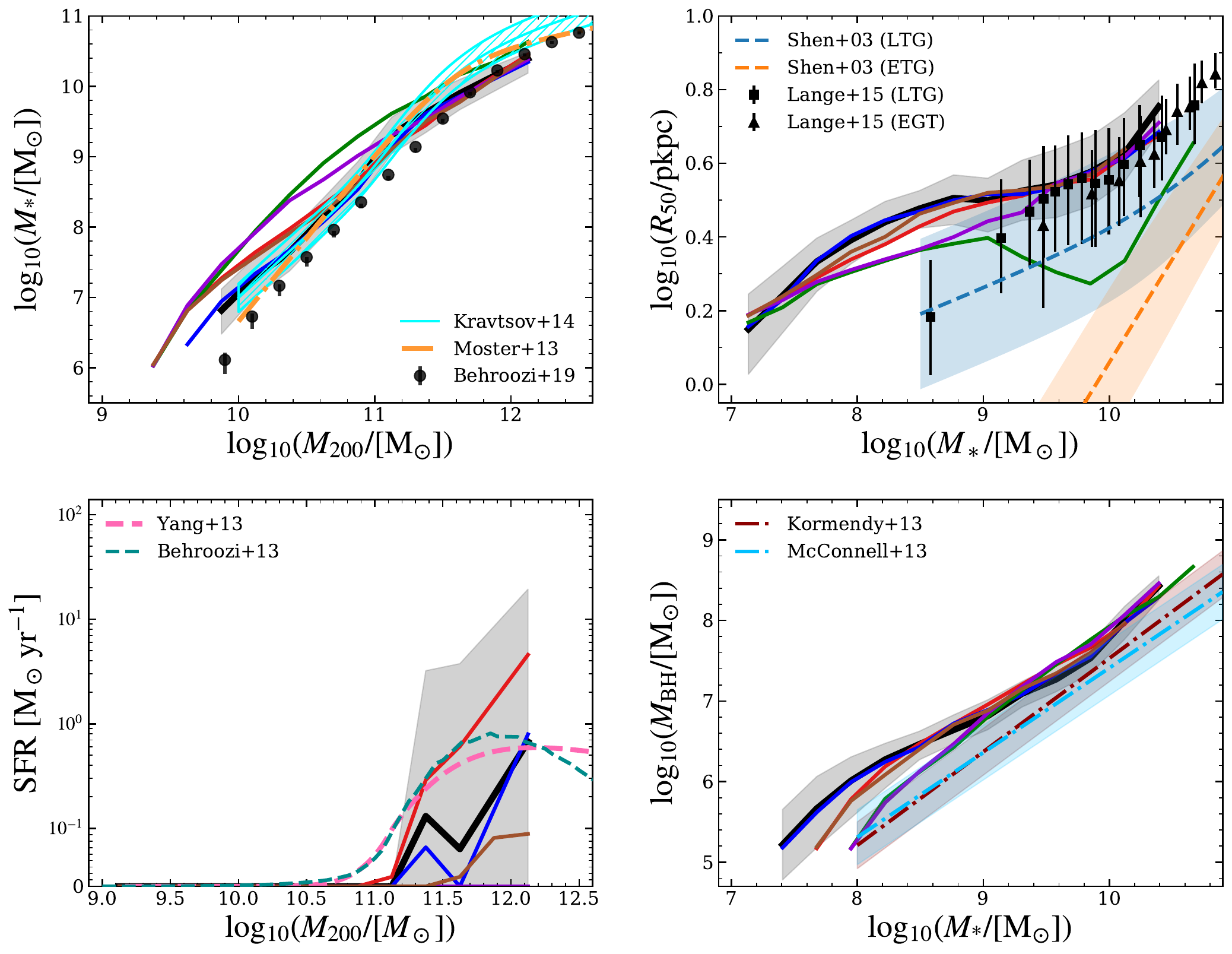}
    \caption{Comparison of the simulated galaxy scaling relations with observational results at $z=0$. The black solid line shows the $\Lambda$CDM results, while the colored lines for the PMF models are colored following the convention in the main text.
    \textbf{Top-left}: The simulated $M_{*}$-$M_{200}$ relation is compared with \citet{2019MNRAS.488.3143B}, \citet{Kravtsov:2014sra}, \citet{moster2013galactic}. \textbf{Top-right}: The simulated $R_{50}$-$M_*$ relation is compared with the local galaxy size-mass relations of late-type and early-type galaxies from \citet{Shen:2003sda} and \citet{lange2015galaxy}. \textbf{Bottom-left}: The simulated SFR-$M_{200}$ relation is compared with \citet{Yang:2013cdx} and \citet{2013ApJ...770...57B}. \textbf{Bottom-right}: The simulated $M_{\rm BH}$--$M_{*}$ relation is shown together with the local observations from \citet{2013ARA&A..51..511K} and \citet{2013ApJ...764..184M}.}
    \label{fig:z0-observation}
\end{figure*}

Fig.~\ref{fig:z0-observation} compares four key scaling relations from our simulations at $z=0$ against observations: stellar mass versus halo mass, half-stellar-mass radius versus stellar mass, star formation rate versus halo mass, and SMBH mass versus stellar mass. These results, together with the stellar mass function presented in Fig.~\ref{fig:smf}, demonstrate that our $\Lambda$CDM simulation achieves reasonable agreement with several key observational constraints.

\clearpage

\input{main.bbl}
\end{document}

%% file: ads_macros.tex
\providecommand{\araa}{Annu. Rev. Astron. Astrophys.}
\providecommand{\apj}{Astrophys. J.}
\providecommand{\apjl}{Astrophys. J. Lett.}

\providecommand{\aap}{Astron. Astrophys.}
\providecommand{\aapr}{Astron. Astrophys. Rev.}

\providecommand{\jcap}{JCAP}

\providecommand{\mnras}{Mon. Not. R. Astron. Soc.}

\providecommand{\ssr}{Space Sci. Rev.}

\providecommand{\prd}{Phys. Rev. D}

\providecommand{\prl}{Phys. Rev. Lett.}
